# Single and Few-Particle States in Core-Shell Nanowire Quantum Dots

M. Khoshnegar[2,3] and A. H. Majedi[1,2,3]

[1]*School of Engineering and Applied Sciences, Harvard University, Cambridge, MA 02138*
[2]*Institute for Quantum Computing,* [3]*Waterloo Institute for Nanotechnology, Waterloo, Ontario, Canada N2L 3G1*

The electronic properties of single and few-particles in core-shell nanowire quantum dots (NWQD) are investigated. By performing configuration interaction (CI) calculations we particularly elucidate how elevated symmetry character ($C_{3v}$ or $D_{2d}$) exhibited by single particle orbitals enhances the phase coherence of exciton-photon wavefunction though suppressing spin flip processes. Detailed calculations presented here demonstrate how strain-induced potentials manipulate the symmetry characters, intrinsic oscillator strength and electron-hole dipole in NWQDs. An orbital-dependent kinetic energy is defined based on single particle dispersion and orbital spreadout in k-space. It is shown the exchange occurring between this kinetic energy and strain-induced potentials is responsible for orbital distortions, and thus the energy reordering of different direct and correlation terms. Various structures have been examined to elaborate on the influence of size and orientation together with axial and lateral symmetry of NWQDs. Our many-body calculations suggest that binding energies of *s*-shell few particle resonances $X_0^{\pm}$ and $XX_0$ are suppressed when axial and lateral localizations become comparable. Then exerting an external perturbation may renormalize the binding energies, realizing a transition from anti-binding to binding regime or reverse. In this regard, we specifically show that kinetic energy of single particles, and thus correlation energies of associated complexes, exposed to an electric field remain relatively unaffected and the interplay between direct Coulomb terms reorders the multiexcitonic resonances. Sub-$\mu$eV fine structure splitting along with the tunable $XX_0$ binding energy offers NWQDs promising for generating entangled photons in both regular and time reordering schemes.

## I. INTRODUCTION

An outstanding goal in quantum communication is the realization of genuine single and entangled photons. Reliable sources of entangled photons are undoubtedly the indispensable ingredients of quantum information circuits [1] [2]. The unique atom-like properties of QDs can be utilized to generate polarization-entangled photons through the sequential recombinations occurring in their excitonic complexes, specifically in the biexciton-exciton ($XX_0$-$X_0$) cascade process [3]. The relatively weak Coulomb interactions, however, prohibit any energy equivalence between the excitonic levels and thus the emitted photons [4]. Particularly in $XX_0$-$X_0$ cascade, the anisotropic long range exchange interaction ruins the color coincidence between the intermediate exciton $X_0$ states, and thus the energy indistinguishability of twin photons, by introducing a fine structure splitting (FSS) into the QD energy diagram [5] [6]. Two alternatives are proposed to preserve the parity of emitted photons. First (regular scheme): eliminating FSS via performing structural manipulations [7] or exerting field-related perturbations [8] [9] [10]. As FSS is firmly linked to the symmetry character of exciton wavefunction, it can be erased through symmetrizing the QD geometry and suppressing the built-in piezoelectric fields [11]. Additionally, electric field Stark effect [12], magnetic field [13], strain modulation [14], coupling to cavity modes [15] and spectral filtering [5] are the most commonly reported approaches successfully removing or hiding the path information in $XX_0$-$X_0$ cascade. Second (time reordering scheme): driving the biexciton $XX_0$ binding energy toward zero in order to establish color coincidence "across generations" [16] [17]. This, however, demands for precisely erasing the time information belonging to $XX_0$ and $X_0$ photons. Smaller $XX_0$ binding energy improves entanglement concurrence although it also depends on $XX_0$ and $X_0$ half-line widths [12].

Number of experiments have recently focused on composition-modulated nanowires for generating single photons with second order cross correlation factors $g^{(2)}$ well below 0.5 [18] [19]. In contrast to self-organized QDs, the modulated segment barely suffers from any morphological anisotropy along the growth direction, thus intrinsically forbids the formation of low symmetry (e.g. $C_{2v}$) piezoelectric confinement. This offers ideal features for producing polarization-entangled photons under the regular scheme because the indistinguishability of photons is well-preserved. Furthermore, direct Coulomb interactions and thus the binding energies of few-particle complexes can be optimally tailored in a field-effect transistor configuration [20]. Entanglement in time reordering scheme then could be improved through $XX_0$ binding energy removal.

While a great deal of theoretical and experimental works focus on self-organized QDs, there is no study on few-particle states in NWQDs. In this work, we first investigate how single particle characteristics of core-shell NWQDs depend on their structural parameters. For this purpose, orbitals symmetry character and band mixing effects are analyzed. The impact of single particle dispersion on its total energy and orbital deformations is discussed. Then a many-body picture is provided to describe the few-particle resonances in the excitonic spectrum of NWQDs. By introducing the orbital dependent kinetic energy and observing its variations in response to the confining potentials, we are able to predict the evolution of direct interactions and correlation energies. This way, we can engineer NWQD structure to achieve small binding energies for s-shell complexes. It is demonstrated here that trivial binding energies are likely when axial quantization does not exceed the lateral one significantly. However, fine structure energies calculated here by CI method for s-shell excitons suggest that spin flip processes are suppressed in NWQDs when axial localization is considerably dominant. Finally, NWQDs are exposed to axial and lateral electric fields so that $XX_0$ binding energy is removed. Based on the trivial variations of kinetic energy, we prove that correlations are rather insensitive to the electric field in NWQDs studied here. Thus s-shell excitonic resonances move according to direct Coulomb interactions.

This paper is organized as follows. In Sec. II we address the issue of fine structure splitting in QDs and detail the regular and time reordering schemes. In Sec. III different NWQD shapes studied in this paper are introduced and the appropriate choices of dimensions are justified. They are predominantly related to the anticipated spin relaxation times and the NWQD ground state HH/LH characters. In Sec. IV we discuss the strain induced potentials, band mixing effects and piezoelectricity. In Sec. V we demonstrate how single particle orbitals, their symmetries and mixings follow the interplay between different terms of confinement. The many-body description of few-particle complexes along with the calculation of direct, exchange and correlation interactions are presented in Sec VI. The influence of electric field on direct Coulomb terms and binding energies are examined in Sec. VII, and a summary of our studies is available in Sec. VIII.

## II. NWQDS AND ENTANGLED PHOTON PAIR GENERATION

### A. Fine Structure of Typical QDs

It is known that the anharmonic set of excitonic energies, developed owing to the relatively weak Coulomb interactions in strong confining QDs, prepares a multilevel system appropriate for emitting anti-bunched photons having distinct and tunable energies [21]. Fig. 1 (left panel) illustrates the excitonic energy level scheme of QD ground state with the inclusion of direct and long range exchange Coulomb interactions, which are responsible for biexciton $XX_0$ binding energy $\delta_b$ and FSS $\delta_s$, respectively (scales are not the same). Two photons emitted in each sequential decay possess opposite circular (identical rectilinear) polarizations and dissimilar energies. Distinguishability of the two decay paths, resulted from FSS, degrades photon pair degree of entanglement, unless exciton linewidths are broad enough or properly dressed to "hide" the $X_0$ energy splitting [22].

FSS appears when the anisotropy of QD structure enforces $X_{0,1}$ and $X_{0,2}$ wavefunctions to extend along non-equivalent lateral crystallographic orientations. This anisotropy, which may exist in either axial or lateral dimensions, reduces the exciton wavefunction symmetry. Lateral anisotropy, mostly arising in [001]-oriented self-assembled QDs rather than NWQDs, is a consequence of different surface mobilities along the in-plane directions, namely [110] and [1$\bar{1}$0] [23]. axial anisotropy, however, depends on QD formation during the growth process and its final shape: any morphological asymmetry along the growth direction gives rise to built-in piezoelectric field which may elongate the exciton wavefunction toward different lateral orientations. There have been a number of approaches proposed to suppress the internal piezoelectric field including thermal annealing [24] and growing QDs on (111) substrates [25]. Accordingly, FSS can be interpreted from the symmetry point of view: symmetry properties dictated by QD confining potential plays the key role in FSS, and QDs lacking $D_{2d}$, $C_{4v}$ or $C_{3v}$ symmetry exhibit this energy splitting [11]. As we will see in the following sections, even in laterally symmetric QDs, internal piezoelectric field lowers the wavefunction symmetry, e.g. down to $C_{2v}$, and introduces energy separation between ground state exciton energy.

Fig. 1 (right panel) shows the energy level scheme when the biexciton binding energy is erased, i.e. $\delta_b = 0$. Since the biexciton level is lined up equal to sum of the intermediate exciton energies, $E_{XX_0} = E_{X_{0;1}} + E_{X_{0;2}}$, first photon in each path has identical energy as the second photon in the other path. Provided this condition is satisfied, photons having similar energies can be polarization-entangled in time reordering scheme under specific timing circumstances [16].

### B. NWQDs Versus Self-assembled QDs

III-V nanowire based QDs are typically hetero-insertions composed of two different band gap materials commonly found in zinc-blende or wurtzite phases. The precisely controlled morphology and structure of QD segments during the growth process allows for fabricating cylindrical or hexagonal QDs [26] [27] [28] which, due to their symmetric geometry, turn out to be the primary competitors of self-assembled QDs for generating entangled photon pairs. Tailoring the number of charge carriers down to one single particle trapped in a single gate-induced QD has been already demonstrated by Björk *et al* [29], opening the same possibility for its heterostructure twins. Moreover, the embedded QD segment covers nearly the entire nanowire cross sectional area, thus compared to self-assembled structure the internal quantum efficiency increases remarkably having the common surface states removed or suppressed by means of appropriate shell passivation [28]. On the other hand, unwanted satellite peaks are predicted to be visible in the absorption spectrum of NWQDs supporting compressional acoustic phonon modes with nonvanishing radial part [30]. The zero phonon line (ZPL) is also expected to be broadened (by $\sim 0.5\ \mu eV\ T/K$) due to the nanowire elongation mode. This broadening, motivated by the dimensionality of hosting media (semi-1D), even exceeds the net linewidth seen in the ZPL of bulk embedded QDs [31], and thus may further restrict the exciton coherence time.

### III. SPIN COHERENCE AND POLARIZATION ANISOTROPY IN NWQDS

In this section we address the other essential figure of merit one NWQD as an entangled photon source should possess. Besides the color coincidence within or cross generations, exciton spin coherence substantially affects the degree of entanglement. We first select NWQD insertions having particular dimensions, then justify these choices based on physical perspectives.

### A. NWQD Shapes, Dimensions and Growth Directions

In this paper, we model zinc-blende core-shell NWQDs, having their symmetry properties inspired by a number of experiments recently done [32] [28] [26] [33] [34]. We particularly investigate the built-in potential and wavefunction symmetries when the growth direction varies from [001] to [111]. As illustrated in Fig. 2 the geometries of QD insertions include:

A: Truncated pyramidal QD. No modeling is presented for this series and the results only give us the required insight to accomplish useful comparisons between self-assembled and NWQDs [9].

B-1 and B-2: cylindrically symmetric NWQDs grown along [001] and [111] orientations.

C-1 and C-2: laterally symmetric but axially asymmetric [001]- and [111]-oriented NWQDs. These cone-shaped zinc-blende QDs qualitatively exhibit similar properties to their wurtzite counterparts originally grown by E. Minot *et al* [32].

D-1 and D-2: [001] and [111]-oriented NWQDs with hexagonal cross section.

F: Laterally elongated [001]-oriented NWQDs.

The diameters of most studied QDs $D_D$ are kept fixed equal to 20 nm (except in a few highlighted cases) for the sake of comparison, while their height $h_D$ varies changing the vertical aspect ratio, i.e. $a_h = h_D/D_D$, between 0.1 and 0.5. The QD/barrier materials are assumed InAs/GaAs for [001]-oriented and InAs/InP for [111]-oriented QDs, comparable to the previously reported experiments [35] [19]. It is noteworthy to mention that the macroscopic symmetry properties of wavefunctions derived here for [111]-oriented zinc-blende QDs are also extensible to their [0001]-oriented wurtzite counterparts. Additionally, we expect analogous values of Coulomb interactions being observed in wurtzite QDs, since the effective orbital behaviors are rather similar. In series A, two monolayers of QD material are added beneath the pyramid island serving as the so called wetting layer.

The variation of QD vertical aspect ratio is selected between 0.1 and 0.5 because of the following reasons: first, we intend to demonstrate the impact of confinement variations in NW where the effective mass of single particles along one particular coordinate (quantization axis) is principal. The primary character of hole ground state $h_0$, and the associated exciton $X_0$, then can be easily identified among heavy hole (HH), light hole (LH) and spin orbit split-off (SO) classifications of the valence bands. This character directly returns a measure of the angular momentum shared by the hole particle in an exciton and, therefore, the polarization of emitted photon after the recombination process. Band mixing between HH and LH excitons caused by lattice deformations also affects the spin relaxation times which will be further detailed in Secs. III B and VI C 4. Second, energy of the primary interband transition firmly relies on the lowest dimension seen in the QD. We aim to keep the corresponding wavelength within the infrared spectrum, being compatible with other optical components if required. Additionally, this range (0.1-0.5) has been frequently explored by the relevant experimental works on NWQDs. Needless to remind that similar transition energies are achievable through higher vertical aspect ratios using a ternary composition of InAs/GaAs or InAs/InP materials as the QD segment.

The lateral elongation of structure, if sufficiently pronounced, is itself a possible origin of FSS in NWQDs since it reduces the symmetry character into $C_{2v}$ even in the absence of a directional piezoelectric field. The net effect of piezoelectricity and elongation can, however, increase or decrease the polarizability of an exciton wavefunction. In this paper, we do not study the lateral elongation extensively, since our results can be easily revised in the presence lateral asymmetry.

### B. Spin Relaxation and Exciton Cross Dephasing

During the decay process of a neutral biexciton toward ground state, a temporary hybrid wavefunction of the emitted photon and the remained exciton, known as the exciton-photon state, develops after the first exciton is recombined and lives until the second exciton lifetime gets terminated [36]:

$$\psi_{\text{ex-ph}} \propto \left| \lambda_1^{XX} X_{\lambda_1} \right\rangle + \left| \lambda_2^{XX} X_{\lambda_2} \right\rangle \qquad (1)$$

where $\lambda_i$ ($i = 1, 2$) represents the polarization state of the photon and $X_\lambda$ symbolically stands for the remained exciton which has to recombine after a certain time interval, emitting another $\lambda_i$-polarized photon (we assumed rectilinear or diagonal basis, otherwise polarization state changes for the second photon in circular basis). This hybrid state eventually ends up with a two-photon state, which its time evolution and degree of entanglement crucially relies on FSS. Any spin flip occurrence amid the existence of intermediate exciton-photon state randomizes not only the relationship between its present and previous phases in course of the time, $\phi_\lambda(t)$ and $\phi_\lambda(t - \tau)$, but also the relationship between the phases of two intermediate wavefunctions belonging to the two different decay paths [37]. This phenomenon, called as cross dephasing, accompanied by spin scattering they reduce the first-order cross-coherence $g_{\lambda_1,\lambda_2}^{(1)}$ of emitted photons [38]. The cross dephasing characteristic time is typically larger than the exciton lifetime, but this ordering may be affected by the QD size and temperature.

Cross dephasing time is determined by the spin longitudinal relaxation time $\tau_1$ of the exciton-bound electron and hole. Two primary mechanisms are responsible for exciton spin flip process depending on the strength of QD confinement [39] [40]. First mechanism originates from both electron-hole exchange interaction $H_{\text{exc}} = -2/3 \Delta_{bd} \mathbf{S} \cdot \mathbf{J}$ and the phonon-induced deformation $H_{\text{ph}}$. Here, $\mathbf{S}$ and $\mathbf{J}$ are electron spin and hole total angular momentum operators, and $\Delta_{bd}$ is the exchange energy splitting between the dark and bright doublets. Expanding on the basis of the bright HH and LH states, $|\pm 1\rangle_{hh}$ and $|\pm 1\rangle_{lh}$, the total hamiltonian reads

$$H_{exc} + H_{ph} = \begin{pmatrix} 0 & 0 & \frac{\Delta_{bd}}{\sqrt{3}} & R^* \\ 0 & 0 & R & \frac{\Delta_{bd}}{\sqrt{3}} \\ \frac{\Delta_{bd}}{\sqrt{3}} & R^* & \Delta_{hh\text{-}lh} & 0 \\ R & \frac{\Delta_{bd}}{\sqrt{3}} & 0 & \Delta_{hh\text{-}lh} \end{pmatrix}. \qquad (2)$$

$\Delta_{hh\text{-}lh}$ represents the energy spacing between HH and LH excitons originated from strain forces. In the case of [001]-oriented nanowires, for instance, in which $x$ and $y$ label the in-plane inequivalent vectors ($x = [100]$ and $y = [010]$), $R$ is $\sqrt{3}/2 \, b_v (e_{xx} - e_{yy}) - i d_v e_{xy}$, where $b_v$ and $d_v$ denote shear deformation potentials along [001] and [111] orientations, respectively, and $e_{ij}$ is the strain tensor component. As described in the multiband picture of zinc-blende crystals (see Sec. IV A), the deformation term $R$ comes from the structural anisotropy and directly contributes to the HH-LH level splitting. Cross dephasing occurs between the two bright exciton states $|X_{B;1,2}\rangle = 1/\sqrt{2} \, (c_\uparrow^\dagger h_\Downarrow^\dagger \pm c_\downarrow^\dagger h_\Uparrow^\dagger) | s_e, s_h \rangle$ (where $c_\sigma^\dagger$ and $h_\sigma^\dagger$ each create an electron and a hole with spin $\sigma$ in ground states $|s_e\rangle$ and $|s_h\rangle$, respectively) that are now energetically separated, but perturbatively mixed with the introduction of anisotropy. Both electron and hole spins are assumed to simultaneously flip in this mechanism. The corresponding matrix element then reads

$$\langle X_{B;1} | H_{\text{exc}} + H_{\text{ph}} | X_{B;2} \rangle \propto \frac{\Delta_{bd}}{\Delta_{hh\text{-}lh}} b_v (e_{xx} - e_{yy}) \qquad (3)$$

Considering only one acoustic-phonon mode among the whole phononic spectrum, the relaxation time inversely relates to the corresponding oscillator strength $|\langle X_{B;1}|H_{\text{exc}}+H_{\text{ph}}|X_{B;2}\rangle|^2$, which means

$$\frac{1}{\tau_{\text{deph}}} \propto \frac{\Delta_{bd}^2}{\Delta_{hh\text{-}lh}^2} b_v^2 (e_{xx} - e_{yy})^2. \quad (4)$$

Eq. (4) states that the rate of spin relaxation in this mechanism depends on QD lateral asymmetry and decreases when the strained HH band is sufficiently away from the nearest LH band. FSS also contributes in the spin flip rate through phonon emission and absorption accompanying this process; i.e. $1/\tau_{\text{deph}} \propto \delta_s^3$ provided that $\delta_s D_D / 2\pi\hbar \leq v_s$, where $v_s$ is the sound velocity [39]. By changing the QD shape or dimensions $\Delta_{bd}$ remains nonzero independent of in-plane symmetry. In small QDs hosting Frenkel excitons (typically $D_D <$ 15 nm), electron-hole short range exchange interaction predominantly causes $\Delta_{bd}$, but long range exchange interaction further contributes as the QD size becomes larger. Conclusively, symmetric QDs (where $|X_{B;1}\rangle$ and $|X_{B;2}\rangle$ are not energetically separated) remain rather safe against simultaneous spin flip process ($|s_{e,\uparrow},s_{h,\Downarrow}\rangle \leftrightarrow |s_{e,\downarrow},s_{h,\Uparrow}\rangle$).

Second mechanism includes an intermediate transition to one dark state (electron and hole spins flip sequentially), where spin-orbit interaction stimulates the process and exchange splitting serves as an external magnetic field. Notice that Dresselhaus spin-orbit interaction permanently exists regardless of the structure inversion symmetry since the underlying zinc-blende crystal lacks any inversion center. This mechanism seems to be dominant in QDs with large axial confinement (small height; e.g. $h_D \ll \{D_D, a_B\}$, where $a_B$ is the exciton Bohr radius) but weak lateral quantization, i.e. $a_h \leq 0.2$ and $m_e^* \sim m_{h\perp}^*$, or large effective diameter e.g. $D_D > 25$ nm. As discussed in the following sections, HH-LH mixing is also negligible for this type of QDs ($\Delta_{hh\text{-}lh} > 10$ meV). For $\Delta_{bd} < 2\pi\hbar v_s/D_D$ the electron (hole) spin flip rate is effectively determined by coupling to the long-wave phonons. Assuming a simplified harmonic potential in the vicinity of Γ point in k-space the spin relaxation rate for exciton-bound electron (hole) then could be approximated as [40]

$$\frac{1}{\tau^{e,h}} \propto \frac{\Delta_{bd}^5}{\hbar^2 \omega_0^2 (\hbar\omega_0 + \Delta_{bd})^2} \left(\frac{m_{e,h\perp}^*}{m_e^* + m_{h\perp}^*}\right)^2 \beta_{e,h}^2 (N_{\text{ph}} + 1) \quad (5)$$

where $\hbar\omega_0$ stands for the lateral quantization energy ($\omega_0 \propto 1/D_D^2$), $\beta_{e(h)}$ denotes the electron (hole) spin-orbit coupling coefficient and $N_{\text{ph}}$ is the phonon occupation factor which prescribes the quantity of available phononic modes. Under the limit $D_D < 2\pi\hbar v_s/\Delta_{bd}$, spin relaxation rate grows in proportion to $\sim \Delta_{bd}^5$ ($\Delta_{bd} \ll \hbar\omega_0$) and $\beta_{e(h)}^2$ when coupled to one phonon mode. The relaxation rates calculated in Sec. VI C 4 suggest that this type of single particle spin flip may even occur during the exciton lifetime (in the order of a few ns or shorter). This Instability of exciton spin configuration becomes particularly important for those materials having noticeable spin-orbit coupling energy; for example InAs with $\Delta_0 = 380$ meV. Above the $D_D \sim 2\pi\hbar v_s/\Delta_{bd}$ limit, contribution from the short-wave phonons partially enhances the spin configuration lifetime.

Above discussion along with the results provided in Sec. VI C 4 brings us to the conclusion that symmetric flat QDs in which the lateral confinement has not been drastically reduced are eligible nominees from the spin coherence point of view. This assumption also justifies our choices of QD dimensions assigned in the previous section, i.e. 2 nm < $h_D$ < 10 nm and 20 nm < $D_D$ < 30 nm.

### C. Polarization Anisotropy in NWQDs

Owing to their high refractive index compared to the surrounding ambient, that is normally air or vacuum, III-V nanowires exhibit extensive polarization anisotropy in their emission and absorption characteristics ($\varepsilon_{\text{static}}$= 13.2 for GaAs and 12.5 for InP, respectively). The component of an external electric field parallel to the nanowire axis remains the same inside and outside, $E_{\|;in} = E_{\|;ex} (\lambda \gg D_D)$, while the perpendicular component weakly penetrates inside $E_{\perp;in} = 2\varepsilon_{\text{vac}}/(\varepsilon_{in} + \varepsilon_{\text{vac}}) E_{\perp;ex}$ ($\varepsilon_{in}$ and $\varepsilon_{\text{vac}}$ represent the nanowire and vacuum dielectric constants). Assuming that $\mathbf{e}_{\mathbf{k}\lambda} \in \{\|,\perp\}$ symbolizes the polarization vector for an electric field possessing wavevector $\mathbf{k}$ and wavelength $\lambda$, the measures of associated Rabi frequency $\chi_{i,f} = \mathbf{p}_{i,f} \cdot \mathbf{E}_{in}$ ($\hbar = 1$) and the effective oscillator strength $M_{i,f}^{\text{eff}} = E_{\mathbf{e}_{\mathbf{k}\lambda};ex}^2 / E_{\mathbf{e}_{\mathbf{k}\lambda};in}^2 M_{i,f}$ then behave very differently in response to parallel and perpendicular polarizations for an interband transition between states $|i\rangle$ and $|f\rangle$ as $E_{\perp;in} \ll E_{\|;in}$ [$2\varepsilon_{\text{vac}}/(\varepsilon_{in} + \varepsilon_{\text{vac}}) \ll 1$] for an arbitrary external electric field unless $E_{\perp;ex} \gg E_{\|;ex}$. The oscillator strength we refer to throughout this paper is the intrinsic one $M_{i,f}$.

The symmetry character of a Bloch spinor corresponding to a specific state introduces another source of anisotropy into the interband transitions, which is not exclusive to NWQDs. With the inclusion of band mixing (see Sec. IV A) and neglecting the envelope symmetries, those states exhibiting larger HH (LH) character have $E_{3/2}$ ($E_{1/2}$) symmetry according to the group notation used in Ref. [41]. Electron spinors possess $E_{1/2}$ symmetry, thus any transition from CB ground state to HH-like states is allowed only under perpendicular polarization, i.e. $\langle c_{E_{1/2}} | \mathbf{E}_\parallel \cdot \mathbf{p} | hh_{E_{3/2}} \rangle = 0$. In contrast, interband transitions from CB to LH-like states sharing a similar symmetry are allowed for both types of polarizations, i.e. $\langle c_{E_{1/2}} | \mathbf{E}_{\parallel,\perp} \cdot \mathbf{p} | lh_{E_{1/2}} \rangle \neq 0$. Consequently, the effective oscillator strength drops when the hole effective mass becomes heavier along the nanowire axis (more HH-like). Having these facts in mind, we discuss how the VB ground state is shared by HH and LH bands of the original bulk in Secs. IV and V.

## IV. STRAIN CONFINEMENT AND PIEZOELECTRICITY

Since the confining potentials caused by strain field are comparable to the original band offsets of III−V compounds, one cannot neglect their influence on the single particle energy levels and the corresponding wavefunctions, especially when strain becomes highly anisotropic. Having employed the envelope function approximation (**k.p** model) to solve for the single particle orbitals, we followed the continuum mechanical model (CM) in order to find the strain distribution [42] [43]. **k.p** hamiltonian is generally insensitive to the complete details given by atomistic models, such as valence force field (VFF) [44] [45]; while with a smaller load of computations, CM can properly fit the restricted number of predefined parameters introducing the strain field into this model [46]. In order to model the strain distribution, we assumed epitaxial growth. A detailed study of calculation method including initial and boundary conditions along with elastic constants can be found in Appendix A and Refs. [6] and [47].

### A. Band Alignment and Wavefunction Coupling

The role of strain in adding extra confinements to the original band alignments of QD/nanowire boils down into three types of potential: 1) hydrostatic $V_{\text{Hy}}$, 2) biaxial $V_{\text{Bi}}$ and 3) piezoelectric $V_p$. Origin of the two first types, $V_{\text{Hy}}$ and $V_{\text{Bi}}$, lies in the diagonal components of strain tensor, i.e. $\text{Tr}\{e\} = e_H = e_{xx} + e_{yy} + e_{zz}$ and $e_B = (e_{xx} + e_{yy})/2 - e_{zz}$, while piezoelectric potential originates from the off-diagonal components (see Sec. IV. C). Let us exemplarily consider the case of [001]-oriented NWQDs. Proper approximations of QD conduction band (CB) and valence band (VB) energy eigenvalues are given by their energy band alignments having the strain effect included:

(i) The local position of conduction band at $\Gamma_6$ neighborhood in k-space merely depends on the hydrostatic strain: $E_{CB} = E_{CB}^0 + a_c e_H$, where $E_{CB}^0$ denotes the strain-free CB edge and $a_c$ is the CB hydrostatic deformation potential. Above relation implies that higher strain energy developed at QD interfaces lifts up electron eigenvalues in the energy space.

(ii) The position of VB edge, however, relies upon both hydrostatic and biaxial potentials. The HH band at $\Gamma_8$ in k-space curves according to $E_{hh} = E_{VB}^0 - a_v e_{Hy} + b_v e_{Bi}$, where $E_{VB}^0$ is the unstrained VB edge, and $a_v$ stands for the valence band hydrostatic deformation potential. In this approximate picture, the energy splitting between HH and LH bands is then $2b_v e_B$ ( $2\sqrt{3}d_v e_{xy}$ in the case of [111]-oriented QDs) [48]. By solving the eight-band **k.p** hamiltonian at $\Gamma$ point, one observes that high axial anisotropy in QD structure yields a local crossing between HH and LH bands leading to the mixture of their associated wavefunctions [46] [49]. This band crossing is attributed to the probable sign-changing of biaxial strain and must be preferably avoided to have purely $\pm 3/2$ ($J_z = \pm 3/2$) or $\pm 1/2$ ($J_z = \pm 1/2$) p-spinors for holes residing the ground state.

In addition to the proximity of HH and LH bands, their coupling ratio depends also on the shear strain through deformation-dependent terms,

$$\langle lh_2 | H_s | hh_1 \rangle = \langle hh_2 | H_s | lh_1 \rangle$$
$$= \frac{\sqrt{3}}{2} b_v (e_{xx} - e_{yy}) - id_v e_{xy} \quad (6a)$$

and

$$\langle hh_2 | H_s | lh_2 \rangle = -\langle lh_1 | H_s | hh_1 \rangle = -d_v (e_{xz} - ie_{yz}) \quad (6b)$$

in the **k.p** hamiltonian $H_{\mathbf{k.p}} = H_k + H_s$ [47], where $H_k$ and $H_s$ indicate kinetic and strain-dependent parts, $|hh_{1,2}\rangle = |3/2, \pm 3/2\rangle$ and $|lh_{1,2}\rangle = |3/2, \pm 1/2\rangle$. As can be realized from above terms, shear strain components together with the QD lateral asymmetry ($e_{xx} - e_{yy}$) give rise to nontrivial mixture of HH and LH wavefunctions. Conclusively, series B-1, B-2 and D-1 flat NWQDs ($a_h \approx 0.1$) are expected to show minimal "impurity" in their hole total angular momentum, thus a single band representation having the maximum contribution from HH band (see Sec. V C for further details); i.e. $|J_h, J_{h;z}\rangle \approx |hh_1\rangle$ or $|hh_2\rangle$. In contrast for series A and C NWQDs, there always exists a non-negligible portion of LH character in their hole angular momentum, i.e.

$$|J_h, J_{h;z}\rangle = \alpha_1 |hh_1\rangle + \beta_{1;\pm} |lh_{1,2}\rangle \quad (7a)$$

or

$$|J_h, J_{h;z}\rangle = \alpha_1 |hh_2\rangle + \beta_{2;\pm} |lh_{1,2}\rangle \quad (7b)$$

where commonly $\beta_{i,\pm} \leq 0.15\alpha_1$, $i \in \{1,2\}$. The average energy separation between HH and LH bands $\sim 2\sqrt{3}d_v e_{xy}$ is relatively small in [111]-oriented laterally symmetric NWQDs (series B-2, C-2 and D-2), and the hole ground-state $|s_h\rangle$ noticeably gains LH character for larger QD heights where $m^*_{h\perp}$ approaches $m^*_{h\parallel}$.

### B. Strain Relaxation Sensitivity on Shell Thickness

Before elaborating on piezoelectricity, let us discuss how the shell thickness (in series B to D) may shift the band alignments. Since we assume that NWQDs are grown in zinc-blende phase, their energy band alignments have been considered to be type I [50]. Compared to VB, the location of CB bottom edge inside QD then sensitively relies upon hydrostatic strain since $a_c \approx 10a_v$ [51]. Consequently, when the surrounding shell is suppressed, the strain energy formed in the vicinity of QD sidewalls is not able to relax properly. In particular, this strain inhomogeneity extraordinarily pushes down the CB at the sidewalls, leading to deep confinement of trapped electrons at QD borders instead of its center. Having the hole density mostly concentrated around the center, electron and hole orbitals spatially separate resulting in a weak oscillator strength. Thicker shells acting like adequately large barriers also prevent fast recombination of electrons through surface states. Above effect is even more pronounced in InAs/GaAs NWQDs compared with their InAs/InP counterparts since the lattice mismatch $e_\perp$ reaches $-6.69\%$, implying a higher compressive force imposed by the majority material. For larger QD insertions, the critical thickness in the vicinity of insertion is enhanced because larger strain energy must be accommodated by the surrounding shell. Choosing appropriate thickness normally exceeds this critical value [52]. In practice, growth of passivating shell usually demands for accurate consideration of critical thickness. In this regard, our models contain experimentally verified measures [53], [28].

### C. Symmetry Degradation Due to Piezoelectricity

Piezoelectricity, known as the accumulation of charge in response to structural deformations, occurs in non-centrosymmetric crystals including zinc-blende or wurtzite phases. Linear piezoelectricity counts exclusively on shear strain components: $\boldsymbol{P}_a = 2d_{14}\ (e_{yz}, e_{xz}, e_{xy})$, where $d_{14}$ denotes the first order piezoelectric moduli [51]. Moreover, one shall not neglect second order piezoelectric field $\boldsymbol{P}_b$ once strain components increase in highly mismatched hetero-structures like In(Ga)As/GaAs. With the aid of *ab initio* calculations, Bester *et al* showed that the magnitude of quadratic term may compensate or even surpass that of linear term in In(Ga)As quantum wells [54]. This situation particularly applies to QDs where a high degree of anisotropy is observed. In addition to the shear strain, quadratic term depends also on diagonal components of strain tensor which are more significant interior the QD. This yields second order piezoelectric field $\boldsymbol{P}_b$ to appear predominantly inside QDs or close to their boundaries. Depending on the QD geometry, this internal field may intensify or suppress the impact of $\boldsymbol{P}_a$.

In the following, we examine different NWQD series introduced in Sec. III A, and establish a relationship between their geometry and their confinement symmetry. We demonstrate how single particle orbitals are distorted under these symmetry conversions. We also highlight how sensitively these orbitals feel the axial confinement, strain and piezoelectricity by providing their expectation values versus size variations. Notice that throughout this paper, our coordinate system is chosen as $x = [100]$, $y = [010]$ and $z = [001]$.

## V. SINGLE PARTICLE ORBIRALS

In this section we study the global symmetry of electron and hole orbitals along with their associated energies in our QDs. The Bloch components of single particle wavefunctions are assumed *s*-like for electrons and *p*-like for holes, respectively [55]. Geometry of the QD, however, manipulates the symmetry character of the envelope part. In addition to the excitonic energy splittings, orbital symmetry determines the intrinsic oscillator strength $M_{i,f}$ [11]. According to the HH-LH band mixing described in Sec. IV A, general selection rules fail to accurately explain the excitonic transitions. However, the magnitude of intrinsic oscillator strength roughly depends on the type of symmetry shared by electron and hole total wavefunctions [56].

In order to find the single particle envelope functions corresponding to $|C\rangle$, $|X\rangle$, $|Y\rangle$ and $|Z\rangle$ Bloch spinors, 8-band **k.p** hamiltonian has been diagonalized [46]. A similar version of this approach under gauge invariant discretization has been previously employed to calculate the effective *g*-factor of non-passivated NWQDs [57]. Details on the calculation of single particle orbitals can be found in Appendix A.

### A. Symmetry Characters of Net Confinement and Orbitals

#### 1. *Series A, self-assembled QDs*

Many theoretical works has already been devoted to study the single particle states in self-assembled QDs.

Structures having drawn more attention in these analyses include pyramidal and lens-shaped QDs [58] [59], with their geometries supporting elevated $C_{4v}$ and $C_{\infty v}$ symmetry characters, respectively. The net confinement, however, may contain lower degrees of symmetry depending on the growth orientation. Pyramidal QDs suffer from an extreme axial anisotropy which may be moderated by truncating their top region. This eventually brings down the symmetry of electron and hole orbitals to $C_{2v}$: in anisotropic [001]-oriented QDs, single particle orbitals tend to extend toward [110] and [1$\bar{1}$0] directions as a result of piezoelectricity, thus the prominent exciton emission lines are seen along these directions. This suggests that the associated bright excitons $X_{B;1,2}$ are polarized along [100] and [010], as the polarization of each emission line is a linear combination of bright exciton polarizations due to the spin-orbit coupling. Any structural defect perturbs the equal weights in this linear combination thus deviates the polarizations from [110] and [1$\bar{1}$0] [60]. A. Schliwa et al have argued that how properly adding annealing steps may restrain this deviation [58] [61]. Analogous to self-assembled pyramids, orbitals choose $C_{2v}$ symmetry in lens-shaped QDs. However, since the geometry is $C_{\infty v}$-symmetric, replacing (001) substrate with (111) lifts up the net confinement symmetry to $C_{3v}$ leading to a vanishing FSS. Apart from growing highly symmetric structures, strain manipulation, electric field and magnetic field (acting like an extra SO interaction [62] [63]) may be applied to remove FSS.

## 2. Series B-1 and B-2 NWQDs

Unstrained series B-1 NWQDs are $D_{\infty h}$ symmetric. Piezoelectric field, however, follows the symmetry imposed by shear strain components ($e_{yz} = C_z^{\pi/2} e_{xz}$, and $P_z(r<D_D) \approx 0$ since $e_{xz}$ is trivial inside QD), leading to $D_{2d}$-symmetric potential; see Fig. 3(a) left panel. Piezoelectric potential is negligible deep inside the QD insertion where the dense parts of electron and hole orbitals commonly exist. Fig. 3(b) left panel, illustrates Fourier transform of the s-shell electron and hole orbitals, $\rho_{e_0} = |\mathcal{F}(|\varphi_e^s(\mathbf{r})|^2)|$ and $\rho_{h_0} = |\mathcal{F}(|\varphi_h^s(\mathbf{r})|^2)|$, along with the differential probability density $\Delta\rho_0 = \rho_{e_0} - \rho_{h_0}$ in k-space. $\rho_{e_0}$, $\rho_{h_0}$ and $\Delta\rho_0$ are shown on (001) plane in the vicinity of $\Gamma$ point (2 nm$^{-1}$ × 2 nm$^{-1}$ large area). $\rho_{e(h)}$ spreadout in k-space represents a measure of electron (hole) density in real space: since $\rho_{h_0}$ is more extended than $\rho_{e_0}$ in k-space we expect its orbital $|\varphi_h^s(\mathbf{r})|^2$ to be relatively denser. $\Delta\rho_0$ is influenced by the piezoelectric effect suppressing e/h spatial overlap, thus the corresponding oscillator strength $M_{|s_e\rangle,|s_h\rangle}$, and its average value determines e/h dipole.

Noticeable portions of orbitals move toward the QD corners due to the net result of first and second piezoelectric potentials. Fig. 4(a) shows e/h orbitals $\Phi_{e,h}^{s,p} = \sum_{m_J} |\varphi_{e,h}^{m_J;s,p}|^2$ ($m_J$ sweeps over the total angular momentum numbers) for the s-shell $|e_0,h_0\rangle$ (ground state) along with the two excited p-shells, $|e_1,h_1\rangle$ and $|e_2,h_2\rangle$. axial confinement ($\Xi_{v;e} = \Delta_{CB} + \langle H_s \rangle_{v;e}$ for electron and $\Xi_{v;h} = \Delta_{VB} + \langle H_s \rangle_{v;h}$ for hole) is tight when $a_h$ gets relatively small, e.g. $\leq 0.2$, preventing flexible orbital deformations. Furthermore, since the piezoelectric potential scales up in proportion to the QD size, the built-in field has not yet penetrated the QD interior considerably: $\langle \Psi_{h_0} | V_p | \Psi_{h_0} \rangle = \langle V_p \rangle_{h_0} = 0.60$ meV and $\langle V_p \rangle_{e_0} = -0.33$ meV. Increasing $a_h$ leads to vertical spreading of electron and hole orbitals. Hole particle then responds sensitively to the confinement variations (comprising piezoelectric effect) because of the higher effective mass $m^*$ it has (the reason is further detailed in the following): $\langle V_p \rangle_{h_0} = 4.37$ meV and $\langle V_p \rangle_{e_0} = -0.74$ meV for $a_h = 0.4$. This behavior can be observed in $|h_0\rangle$ orbital, where [110] and [1$\bar{1}$0] nodes are not aligned vertically (likewise, $|h_1\rangle$ and $|h_2\rangle$ orbital lobes at [110] and [1$\bar{1}$0] corners are not in the same level).

In series B-2 NWQDs, all three shear strain components show similar patterns under 120° in-plane rotations, i.e. $C_z^{2\pi/3} e_{ij} = e_{i\pm 1\, j\pm 1}$ where $i,j \in \{x, y, z\}$. Same symmetry character, i.e. $C_{3v}$, belongs to the piezoelectric potential as can be seen in Fig. 3(a) right panel. Different components of piezoelectric field then add up to build a [111]-directed field: the oppositely signed isosurfaces at the top and bottom interfaces in Fig. 3(a) right panel ensure the presence of a non-negligible axial electric field. This axial field may suppress $M_{|s_e\rangle,|s_h\rangle}$ by splitting the electron and hole orbitals. Having considered InP hosting nanowire, the amplitude of piezoelectric potential is not comparable to that of In(Ga)As/GaAs hetero-insertions. Hence, the internal axial field in [111]-oriented InAs/GaAs NWQDs becomes remarkably stronger when Indium concentration is large. Nevertheless, $\Delta\rho_0$ in series B-2 ([111]-oriented InAs/InP) is comparable to series B-1 ([001]-oriented InAs/GaAs), implying equivalent e/h dipole (see Fig. 3(b)).

$D_{\infty h}$ symmetry of QD insertion in series B-2 translates to the conformity of orbital symmetry and that of strain-induced potentials. Our calculations show that the net confinement experienced by e/h has $C_{3v}$ symmetry in s- and p-states. When $a_h$ is increased $0.2 \rightarrow 0.4$, $\Xi_{v;e,h}$ relaxes and piezoelectric potentials around the top and bottom interfaces increase. The single particle orbitals then respond freely to the built-in electric field by moving upward or downward, as illustrated in the lower panel of Fig. 4(b). According to the group theory, we expect at least one pair of degenerate bright states ($|\pm 1\rangle_{hh}$ or $|\pm 1\rangle_{lh}$) having pure and isotropic in-plane polarization for the ground state excitonic transition.

Before proceeding further, let us clarify why electron and hole adapt differently to the geometrical variations and potential distortions. Owing to the small effective mass of electron $m_e^*$, its average dispersion $E(\mathbf{k})$ is sharply curved in $\mathbf{k}$-space leading to an abrupt change of kinetic energy in response to any small variation of wavefunction in the real space. Thus, electron orbital is rather resistant to any potential deformation. Hole orbital, on the other hand, accepts any trivial detail of the confinement to keep its potential energy minimized (dissimilarity between electron and hole behaviors is further detailed in Sec. VI C 2). This can be clearly observed in $|h_0\rangle$ orbitals, where the main character of wavefunction comes from HH band: even for $a_h = 0.2$, $|h_0\rangle$ orbital has been to some extent shifted toward the top interface where negative piezoelectric potential exists; see Fig. 3 (a). Further enlargement in [111] direction lifts the hole orbital remote from the electron orbital since a nearly uniform electric field $P_z$ exists in this region. The expectation values given in Table II affirm the dissimilarity of electron and hole $s$-orbitals: once $a_h$ increases, $\langle V_p \rangle_{h_0}$ grows because $|h_0\rangle$ orbital immediately attempts to save its spatial overlap with negative piezoelectric potential. Conversely, $\langle V_p \rangle_{e_0}$ gets suppressed despite the increase of $V_p$ as $|e_0\rangle$ orbital follows the positive piezoelectric potential very slowly.

### 3. Series C-1 and D-1 NWQDs

Piezoelectric potential is shown for series C-1 in Fig 5(a) left panel. Irrespective of their opposite signs, piezoelectric isosurfaces at the top interface become slightly different from their bottom counterparts given larger $a_h$. Although the discrepancy of the top and bottom interfaces results in a global $C_{2v}$ symmetry, but as long as the ratio between top and bottom facet areas is not considerably below 1, e.g. $\geq 0.8$, one can still assign this series a $\sim D_{2d}$ symmetry. This Symmetry resduction would occur in envelope functions and thus $\rho_{e(h)}$. Fig. 5(b) illustrates $\rho_{e_0}$, $\rho_{h_0}$ and $\Delta\rho_0$ on (001) plane (2 nm$^{-1}$ × 2 nm$^{-1}$ large area) in k-space. $\rho_{e_0}$ ($\rho_{h_0}$) exhibit $C_{2v}$ symmetry polarized along [1$\bar{1}$0]([110]). This polarization is more pronounced for $\rho_{h_0}$ due to hole's larger effective mass. Due to the same reason $\rho_{h_0}$ undergoes further deformation when confinement relaxes ($a_h = 0.2 \to 0.4$). Moreover, rapid growth of $\Delta\rho_0$ implies $e/h$ dipole enhancement in response to strain-induced potentials.

Fig. 6(a) depicts the $s$- and $p$-shell electron and hole orbitals in series C-1 NWQDs, where the slight axial anisotropy introduced to the geometry breaks the $D_{2d}$ symmetry down to $C_{2v}$: hole and electron orbitals extend along [110] and [1$\bar{1}$0], respectively. Since the polarization of exciton follows the symmetry character of electron and hole orbitals, we expect the appearance of FSS in this case. As implied form the right panel ($a_h = 0.4$), QD height shall be maintaned small in order to avoid further polarization.

During the modelings we observed that the first and second order piezoelectric potentials have opposite signs in these series of NWQDs, thus attempting to cancel out each other. However, linear term heavily dominated the quadratic term in contrast to the full pyramid or lens-shaped QDs where the second order competes with the first order or even rules the total piezoelectric polarization [58]. In Series B-1, enlarging $h_D$ cannot change the ordering between the two terms. Conversely for series C-1 NWQDs, enhancing $a_h$ increases the shear strain besides its overlap with strain diagonal components, leading to a stronger quadratic term. Table I shows the calculated expectation values of strain-induced potentials ($\langle V_p \rangle$, $\langle V_{Hy} \rangle$ and $\langle V_{Bi} \rangle$) for series B-1 and C-1 NWQDs. Both hydrostatic and biaxial deformation potentials experienced by the $e_0/h_0$ orbitals diminish via enlargeing the QD size, leading to 1) higher excitonic band gaps and 2) smaller HH-LH energy splitting.

We also examined hexagonal NWQDs grown along [001] direction having (110) facets (series D-1). The origin of symmetry reduction here lies in the orientation of QD insertion side facets which cannot be mapped simultaneously on (100), (010), (110) or (1$\bar{1}$0) planes: inconsistency between the number of piezoelectric lobes and the number of side facets ruins the balance of potential weight along the facets. This in turn results in a competition between the symmetry of geometry and that of strain-induced confinement. If the latter is sufficiently strong, as in our case, the global symmetry drops to $C_{2v}$; see Figs. 5(a) right panel and 8(a). For both C-1 and D-1 geometries, the axial gradient of piezoelectric potential at small $a_h$ is not strong enough to noticeably split electron and hole orbitals.

### 4. Series C-2 and D-2 NWQDs

Looking at their axial anisotropy, cone-shaped NWQDs are the moderated version of their lens-like twins with $C_{\infty v}$ symmetry. In lens-shaped QDs, however, the top interface smoothly curves providing distinct boundary conditions for the strain distribution and piezoelectric potential. Needless to mention We note that this macroscopic view carelessly ignores the possible defects or dissimilarities in the atomic scale which may distinguish the top and bottom interfaces even in isotropic geometries [64]. We emphasize again that CM model essentially neglects these kinds of imperfections, where an intrinsic but hidden source of FSS lies in the structure. Nevertheless, the conclusions given here for series C-2 NWQDs are extensible to self-assmbled lens-shaped QDs at least from the viewpoint of confinement symmetry.

As could be understood from above discussions regarding the previous geometries, one of the principal templates resolving the confinement symmetry is the NWQD growth direction. Consequently, similar to series B-2 NWQDs, we predict a $C_{3v}$-like symmetry for series C-2 and D-2. This can be easily deduced looking at their piezoelectric isosurfaces in Fig. 7(a). Appearance of oppsitely-signed potentials at both sides reflects a [111]-oriented possibly dropping $M_{|s_e\rangle,|s_h\rangle}$. This drwaback seems to be a unique feature of QDs grown on (111) substrate although it is partially moderated in cone-shaped geometries: $\Delta \rho_0$ gets suppressed when $a_h = 0.4$ as shown in Fig 7(b). In fact, tendency of the hole orbital to cover the entire details of confinement causes a significant part of it to leak toward the bottom side of NWQD despite the presence of up-pushing electric field.

$C_{3v}$ symmetry of the hole orbitals in both $s$- and $p$-shells indicates how precisely they follow the potential spreadout; see Fig. 6(b). For electrons, however, only the ground state ($s$-shell) represents an elevated $C_{3v}$ symmetry, while $p$-shell orbitals have mainly $C_{2v}$ character especially when QD height increases. This behavior can be attributed to the electron's reluctance of accepting the small contribution reserved by piezoelectric effect in the total confinement. The hydrostatic potential $V_{Hy}$ also exhibits almost a constant value over the NWQD interior with tiny variations adjacent to the borders. Although these variations provide it with a $C_{3v}$ character, only dilute parts of electron orbital feel this symmetry while the denser parts see the constant landscape thus primarily obey the geometry.

In contrast to series C-2, hexagonal NWQDs benefit from the elevated $D_{6h}$ symmetry in their geometry. Once being embedded in dislocation-free $\langle 111\rangle$A- or $\langle 111\rangle$B-oriented hexagonal nanowires, their envelope function symmetry shall decrease in respect to the strain distribution with $C_{3v}$ symmetry; see Fig. 7(a) right panel. Same criterion applies to triangular QDs with elevated $D_{3h}$ symmetry [65] [66]. Consistent with experimentally grown hexagonal nanowires on (111)A substrates, we assumed their cross section to have a $(11\bar{2})$ side facet. Even if the cross secation rotates in (111) plane, the net symmetry still remains $C_{3v}$ because rotating the side facets with respect to nanowire axis does not break it on account of the sixfold in-plane symmetry ($D_{6h}$) that geometry has.

In Figs. 8 (b) and (c), orbitals of electrons and holes resting in $s$- and $p$-shell are illustrated for two different $a_h$, 0.2 and 0.4. Owing to the $D_{6h}$ symmetry of the geometry, $a_h$ cannot manipulate the orbital symmetry but its corresponding density. According to Table II and Fig. 7(b), piezoelectricity has the minor effect among the strain-induced potentials especially close to the NWQD center where the most part of $s$-shell probability density lives. Moreover, hydrostatic potential has a nearly constant profile far inside QDs similar to series B-2 and C-2. Hence, the electron $s$-shell orbital is less affected by the strain-induced symmetry. On the other hand, electron $p$-state orbitals tend to swell near the QD borders, where they follow the $C_{3v}$ symmetry character governed by the strain components. Their corresponding probability densities, $\Phi_{e_1}$ and $\Phi_{e_2}$, also take analogous localizations, which itself implies the trivial shear strain seen by them.

## 5. Series F NWQDs

Any lateral elongation imposed on NWQD structure even more complicates the interplay between the potentials contributing in the total confinement. Basically in series B and C, the $D_{\infty h}$ and $C_{\infty v}$ symmetries of geometry reduce to $D_{2h}$ and $C_{2v}$. When piezo-electricity comes to play, the single particle wave-functions are already polarized, thus any elongation along each arbitrary in-plane axis pronouncedly disturbs the exciton net polarization. The total outcome is then summarized to how these two effects might support or cancel out each other.

Here we analyze the impact of elongation on series B-1. The overal description is then credible to some extent for the other series. Elongation puts no distinction between the two orthogonal in-plane directions [110] and [1$\bar{1}$0] since QD is $D_{\infty h}$-symmetric. We elongated the QD insertion 50% in [100] orientation. There is no experimental evidence reporting such a large amount of lateral elongation [26] [28] [32] [33] [34]. We chose this uncommon value to demonstrate the ultimate deviation of orbitals from their usual spreadout. As illustrated in Fig. 9(a), $s$-shell orbitals extend along the axis of elongation giving rise to excitons polarized in the same direction. Since the original piezoelectric potential is $D_{2d}$-symmetric, $s$-states have been predominantly influenced by lateral relaxation rather than piezoelectric field. In the case of $p$-states, a trade-off between these two arises as $a_h$, and thus piezielectric potential near top and bottom interfaces, increase. It is known that degeneracy of *p*-state $|e_{i=1,2}\rangle$ energies for the electrons relies on the polarization of the corresponding orbitals and thus the underlying confinement [67]. As we will see in the next section, *p*-state energy splitting notably diminishes for highly symmetric NWQDs due to the absence of sizable shear strain. Fig. 9(b) shows that electron *p*-states are initially degenerate, but the degeneracy is lifted up considerably after prolonging the QD. The energy separation becomes larger for $h_D/D_1 = 0.4$. We ascribe this to different measures of $\Xi_{v;e,h}$ present in the two QDs. Electron states are less confined in the second QD, thus the related *p*-shell orbitals freely reshape deviating further from their counterparts.

### B. Single Particle Energies

In the previous section we demonstrated that as long as the morphological symmetry with respect to the growth axis is preserved, vertical aspect ratio cannot break the orbital symmetry. It, however, easily manipulates the single particle localization energy. In this regard, two major processes are affected by $a_h$. First one is the interplay between hydrostatic and biaxial strains. Consider series B-1 NWQDs. Biaxial strain deep inside the QD is negligible provided that the diameter and height are almost equal ($a_h = 1$). Hydrostatic strain is negative due to the compressing forces exerted by the surrounding nanowire and shell. Decreasing QD height perturbs the balance established between the in-plane strain components ($e_{xx}$, $e_{yy}$) and the perpendicular component ($e_{zz}$). Enhanced lateral forces then make the QD insertion to spread axially, leading to a relative dominance of $e_{zz}$ component over $e_{xx}$ and $e_{yy}$. This gives rise to a measurable negative biaxial strain separating the HH-LH bands (notice that $H_s^{hh} = -a_v e_{Hy} + b_v e_{Bi}$, $H_s^{lh} = -a_v e_{Hy} - b_v e_{Bi}$, $a_v > 0$ and $b_v < 0$). Accordingly, uniform increase of biaxial strain over the QD space lifts up HH band yielding a local increase of hole ground state energy $E_0^h$. Meanwhile, the absolute value of hydrostatic strain grows and pushes the hole energies upward. The overall impact of strain on the single particle energies can be seen in Fig. 10: the upward shift observed in hole energies moving from $a_h = 0.5$ down to 0.2 returns to the increase in the absolute values of hydrostatic and biaxial strains. Since both $\langle V_{Hy}\rangle_{h_0}$ and $\langle V_{Bi}\rangle_{h_0}$ evolve monotonously versus $a_h$ (see Table I), the non-monotonous change of hole energies is attributed to the second process.

Second process is the axial confinement experienced by each particle along the quantization axis which becomes important for lower $a_h$. Development of biaxial strain $e_{Bi}$ during the reduction of $a_h$ suppresses the LH fraction participating in the hole *s*-state. This results in a higher effective mass $m^*_{h;z}$ felt by holes residing the ground state (either $\sigma = \Uparrow$ or $\Downarrow$). The corresponding orbital, therefore, adapts more sensitively to the confinement profile. When the $a_h$ drops below ~ 0.2, $\Xi_{v;h}$ compresses the hole orbital. Its large effective mass provides it a semi-flat dispersion around $\Gamma_8$ and allows it remain inside QD accumulating further kinetic energy (see Sec. VI C 2). Due to the same reason, hole orbital cannot swell out into the nanowire core and gain potential energy. Considering that both $|\langle V_{Hy}\rangle|_{h_0}$ and $|\langle V_{Bi}\rangle|_{h_0}$ are ascending for $a_h = 0.2 \to 0.1$, we ascribe the ~ 30 meV redshift seen in the hole energies to the second process.

For the entire range shown in Fig. 10, electron energies experience a blueshift moving backward ($a_h = 0.5 \to 0.1$), because: 1) $a_c \gg a_v$; hence hydrostatic strain affects electron localization energies significantly. 2) Electron orbital shrinks while QD height approaches smaller values raising its kinetic energy. For $a_h$ lower than ~ 0.2, electron orbital diffuses into the nanowire and accumulate potential energy.

An inverse situation occurs when QD height remains fixed while its diameter extends; see Fig. 10(f) where the energy levels are plotted versus the QD diameter. Two types of QDs are examined: series B-1 and F. Irrespective of the energy shifts happening to the electron and hole states, 1D elongation in series F ($C_{2v}$ symmetric) is responsible for the splittings between *p* and *d* shells. Blueshifts are saturated for the hole states as [100]-oriented effective mass $m^*_{h;x}$ (assuming [100]-directed elongation) rapidly gets lighter, thus the hole orbital becomes insensitive to any further size variation in this direction and stops exchanging its kinetic energy.

As can be deduced from Fig. 10, single particle energies in all the highly symmetric NWQDs exhibit a similar behavior versus a wide range of variations in $a_h$. *P*-state splittings do not exceed 1 meV in series B-1, B-2 and D-2. Even in series C-1 and C-2, splittings are small particularly for flat QDs where shear strain is trivial. They, however, strongly depend on the angle between the growth axis and the QD side facets, as it determines the strength of axial anisotropy. In addition, energetic separation of the ground and excited states in the valence band reflects the measure of effective mass: the smaller the splitting, the lager the effective mass. Consequently, effective masses along the quantization axes, $m^*_{[001]}$ and $m^*_{[111]}$, are almost equivalent as no serious distinction can be made between the VB splittings seen in all plots.

### C. HH-LH and HH-SO Band Mixing

As described in Sec. IV A, band mixing is a direct consequence of lateral or axial anisotropy. The strength of wavefunction coupling and state *impurity* has been well-studied for pyramidal QDs in both theoretical and experimental works. Results given by K. Karlsson *et al*, based on their polarization-sensitive micro-PL measurements on In(Ga)As/Al(Ga)As QDs, indicate that a relatively pure character exists in the upmost valence bound states [68]. Y. Niquet *et al* theoretically showed that this purity seemingly relies upon $a_h$ [69]. Their results demonstrate that extending $a_h$ increases the LH percentage in the ground state. This can be explained through the reduction occurring in biaxial strain and relaxation of $\Xi_{v;h}$, which makes in-plane and perpendicular quantizations comparable, thus ending up to higher LH character obtained by the hole ground state. We expect that the same rule applies to series C-2 and D-2 QDs as they respond analogously to both strain-induced and $\Xi_{v;h}$.

Fig. 11(a) depicts HH and LH contributions in the ground state wavefunction of series B-1, C-1 and F NWQDs versus $a_h$. In series B-1, the angular momentum quantization axis is [001]. The hole orbital expectedly redistributes when $\Xi_{v;h}$ slightly quenches. Meanwhile HH-LH splitting tends to vanish as the biaxial strain relaxes. Both effects result in higher LH percentage in the *s*-shell wavefunction. Qualitatively, similar phenomena occur in series C-1 NWQDs except that shear strain more effectively takes part in the band mixing. According to Sec. IV A, the presence of nontrivial shear strain then justifies higher LH fractions observed for series C-1 comparing to series B-1 NWQDs. The impact of lateral elongation on HH/LH proportion is also examined. HH percentage remains fixed for a 50% additional elongation in [100] ([010]) direction, while LH fraction decreases. This reduction translates to larger, although small, SO character gained by the bound state. The weak response of hole ground state can be attributed to its large effective mass $m^*_{[001]}$: for this range of $a_h$, $\Xi_{v;h}$ is dominant. Hole particle then becomes lighter along the in-plane directions, thus hesitating to adapt the lateral variations of geometry.

SO band is energetically away from the ground state (in the order of $\Delta_0$), thus weakly participates in *s*-shell envelope function. Fig 11(b) illustrates the contribution of each character (HH, LH and SO) in the *s*-shell non-excitonic oscillator strength $M^{so}_{|s_e\rangle,|s_h\rangle}$. Although $M^{so}_{|s_e\rangle,|s_h\rangle}$ relies also on $|e_0\rangle$ orbital and SO orbital overlap, its very small amplitude confirms that $|\varphi^s_{h;so}|^2 \ll |\varphi^s_{h;hh}|^2$ and $|\varphi^s_{h;lh}|^2$. SO character generally grows with $a_h$ as can be understood from the single particle hamiltonian where HH-SO coupling is a direct function of shear strain: $\langle hh_2|H_s|so_2\rangle = -\langle so_1|H_s|hh_1\rangle = \sqrt{3}/2\, b_v(e_{xx}-e_{yy}) - i\sqrt{2}\, d_v e_{xy}$ and $\langle so_2|H_s|hh_1\rangle = \langle hh_2|H_s|so_1\rangle = -1/\sqrt{2}\, d_v(e_{xz}-ie_{yz})$. Accordingly, we observe larger SO character in cone-shaped QDs. SO orbital becomes non-negligible in *p*-states where energy spacing to SO band is reduced; see Fig. 11(c).

In the end of this section, we summarized our analysis on NWQDs in Table III along with data for self-assembled QDs compiled form relevant literature. In addition to the symmetry characters, electron *p*-state splittings and ground state HH/LH percentages have been compared. These data imply that improving the degree of symmetry and keeping the NWQDs relatively flat suppress FSS, LH band contribution in the ground state, and electron *p*-state splitting.

### VI. Few Particle States

A serious challenge in the modeling of low dimensional systems comes from the fact that their energy level structure depends on the carrier population of each level. Single particle description is credible for empty QDs filled up maximally with one charge carrier. Immediately after entering the subsequent particles into the picture, three major effects including 1) direct Coulomb interactions, 2) exchange effects and 3) correlations must be considered [70]. The weighting factor between these interactions then depends on the type of few-particle states and the QD structural properties. In strongly confining QDs, the few-particle interactions introduce a perturbation to the total confinement as long as the excitons are presumed to be weakly correlated. As we will see in the following, direct Coulomb interactions play the essential role in ordering the few-particle resonances in our series of NWQDs.

In the rest of this work we study the behavior of ground state exciton $X_0$ (*s*-shell exciton), positively and negatively charged excitons $X_0^\pm$ and biexciton $XX_0$ in different series of NWQDs introduced above. Our primary concern is to compare the binding energies of these few-particle states under geometrical variations of QD insertion and external perturbations. This provides necessary information on how external fields might be applied to appropriately control the spectral resonances in the regular or time reordering schemes.

### A. Configuration Interaction Approach

A reliable approach to cover the correlation effects is expanding the true wavefunctions of few-particle states on a set of Slater determinants rather than a single one, as proposed in modified Hartree Fock (HF) approximation. A linear combination of these Slater determinants then would be a refined guess for each actual wavefunction. This combination includes all possible configurations that can be constructed from the available single particle states, here provided by **k.p** band structure. In our investigated NWQDs, the average dielectric constant and dimensions are adequately large so that excitons are considered Wannier type [71]. We employed the configuration interaction (CI) method to calculate the state mixings and orbital deformations [6]. Some important considerations on implementing this method are provided in Appendix B.

## B. Binding Energies of Ground State $X^+$, $X^-$ and $XX$

The general hamiltonian for a multiexciton complex, describing the many-body interactions, is written as [72]

$$H_{nX^m} = \sum_i E_i^e c_i^\dagger c_i - \sum_i E_i^h h_i^\dagger h_i + \frac{1}{2}\sum_{ijkl} q_{ijkl}^{d,ee} c_i^\dagger c_j^\dagger c_k c_l$$
$$+ \frac{1}{2}\sum_{ijkl} q_{ijkl}^{d,hh} h_i^\dagger h_j^\dagger h_k h_l - \sum_{ijkl} q_{ijkl}^{d,eh} h_i^\dagger c_j^\dagger c_k h_l \quad (8)$$
$$+ \sum_{ijkl} q_{ijkl}^{exc,eh} h_i^\dagger c_j^\dagger c_k h_l.$$

Here, $nX^m$ symbolically stands for a few-particle complex made of $n$ exciton ($X$) or trion ($X^+$ or $X^-$) or alternatively their possible combinations. Summation indexes $i$ ($l$) and $j$ ($k$) contain all the information related to the single particle states (spin included). $h^\dagger$ ($c^\dagger$) denotes the creation operator for holes (electrons). $E^e$ ($E^h$), $q^d$ and $q^{exc}$ represent electron (hole) energy, direct and exchange Coulomb interactions, respectively. After diagonalization, the correlation part associated with the off-diagonal terms reveals in the total energy of the excitonic complex. Following some mathematical simplifications, one can readily deduce from above general hamiltonian that the energy separations between ground-state complexes, i.e. their binding energies with respect to exciton

$$E(X_0) = E_0^e - E_0^h + q_{00}^{d,eh} + q^{corr}(X_0) + q^{exc}(X_0), \quad (9)$$

read [73]

$$\delta(XX_0) = -q_{00}^{d,ee} - q_{00}^{d,hh} - 2q_{00}^{d,eh} + 2q^{corr}(X_0)$$
$$- q^{corr}(XX_0) + 2q^{exc}(X_0) - q^{exc}(XX_0), \quad (10a)$$

$$\delta(X_0^+) = -q_{00}^{d,hh} - q_{00}^{d,eh} + q^{corr}(X_0) - q^{corr}(X^+)$$
$$+ q^{exc}(X_0) - q^{exc}(X^+), \quad (10b)$$

and

$$\delta(X_0^-) = -q_{00}^{d,ee} - q_{00}^{d,eh} + q^{corr}(X_0) - q^{corr}(X^-)$$
$$+ q^{exc}(X_0) - q^{exc}(X^-). \quad (10c)$$

Here, $q^{corr}(nX^m)$ is the total correlation energy of a particular complex $nX^m$. Configurations associated with above complexes include: two configurations for each bright or dark exciton ($c^\dagger_{\uparrow,\downarrow}h^\dagger_{\Uparrow,\Downarrow}|s_e,s_h\rangle$), two configurations for positive or negative trion ($X_0^+$: $c^\dagger_{\uparrow,\downarrow}h^\dagger_{\Uparrow}h^\dagger_{\Downarrow}|s_e,s_h\rangle$ and $X_0^-$: $c^\dagger_{\uparrow}c^\dagger_{\downarrow}h^\dagger_{\Uparrow,\Downarrow}|s_e,s_h\rangle$), and one configuration for biexciton ($c^\dagger_{\uparrow}c^\dagger_{\downarrow}h^\dagger_{\Uparrow}h^\dagger_{\Downarrow}|s_e,s_h\rangle$). More variety of configurations was possible if higher shells could be occupied. In these complexes, the spatial overlap between each two orbitals determines the Coulombic interaction between the two single particles. For an $s$-shell exciton we have

$$q_{00}^{d,eh} = \left\langle \psi_{0;\uparrow,\downarrow}^e(\mathbf{r}_e)\psi_{0;\Uparrow,\Downarrow}^h(\mathbf{r}_h) \middle| C \middle| \psi_{0;\uparrow,\downarrow}^e(\mathbf{r}_e)\psi_{0;\Uparrow,\Downarrow}^h(\mathbf{r}_h) \right\rangle$$
$$= e\left\langle \psi_{0;\Uparrow,\Downarrow}^h(\mathbf{r}_h) \middle| V_{0;\uparrow,\downarrow}^e(\mathbf{r}_e) \middle| \psi_{0;\Uparrow,\Downarrow}^h(\mathbf{r}_h) \right\rangle, \quad (11)$$

where $C = e^2/4\pi\varepsilon_r\varepsilon_0|r_e - r_h|$ and $V_{0;\uparrow,\downarrow}^e$ is the mean-field potential established by $s$-shell electron $e_{0;\uparrow}$ or $e_{0;\downarrow}$. Both spin degrees of freedom are acceptable since the measure of direct interaction disregards spin configuration, $\{\uparrow,\downarrow\}$ for electrons or $\{\Uparrow,\Downarrow\}$ for holes. Similarly,

$$q_{00}^{d,hh} = e\left\langle \psi_{0;\Uparrow}^h(\mathbf{r}_h) \middle| V_{0;\Downarrow}^h(\mathbf{r}_h) \middle| \psi_{0;\Uparrow}^h(\mathbf{r}_h) \right\rangle, \quad (12a)$$

and

$$q_{00}^{d,ee} = e\left\langle \psi_{0;\uparrow}^e(\mathbf{r}_e) \middle| V_{0;\downarrow}^e(\mathbf{r}_e) \middle| \psi_{0;\uparrow}^e(\mathbf{r}_e) \right\rangle, \quad (12b)$$

where Pauli exclusion principle restricts the spin configuration of identical particles occupying the same state. Energy ordering of direct Coulomb interactions falls within four common situations, provided that we stick to the ground state configurations:

(i) Electron and hole probability densities are spatially separated, supressing the attractive Coulomb interaction $q_{00}^{d,eh}$. This situation mostly occurs in QDs suffering from (exposed to) internal (external) electric fields. For instance, [111]-oriented and laterally elongated NWQDs exhibit this feature. (i-a) Hole possesses a denser orbital, thus $q_{00}^{d,hh}$ becomes larger than $q_{00}^{d,ee}$ and $q_{00}^{d,eh}$. Strong piezoelectric field or small $a_h$, tightening the hole localization, may develop this situation. (i-b) Electron possesses a denser orbital making $q_{00}^{d,ee}$ the largest energy value among direct interactions. This situation may occur in highly anisotropic and relatively large QDs, where hole orbital extends to cover the negatively charged regions while electron orbital remains rather unaffected. Relevant example of this case would be series C NWQD when $a_h$ is sufficiently increased ($a_h \geq 0.4$). Neglecting the correlation energies, trion spectral lines fall in between those of bright excitons ($X_{0;B}$: $c^\dagger_\uparrow h^\dagger_\Downarrow |s_e,s_h\rangle$ or $c^\dagger_\downarrow h^\dagger_\Uparrow |s_e,s_h\rangle$) and biexciton ($XX_0$: $c^\dagger_\uparrow c^\dagger_\downarrow h^\dagger_\Uparrow h^\dagger_\Downarrow |s_e,s_h\rangle$).

(ii) Next situation regularly occurs when the spatial overlap of electron and hole orbitals is large enough so that $q_{00}^{d,eh}$ exceeds $q_{00}^{d,ee}$ or $q_{00}^{d,hh}$. (ii-a) Hole probability density encompasses electron orbital. This situation is likely to take place in [001]-oriented NWQDs having large $a_h$, where the size of hole envelope function becomes dominant. (ii-b) Hole orbital gets denser compared to the electron orbital when $\Xi_{v;h}$ is enhanced, leading to a significant hole-hole interaction. In contrast to the case (i), biexciton ($c^\dagger_\uparrow c^\dagger_\downarrow h^\dagger_\Uparrow h^\dagger_\Downarrow |s_e,s_h\rangle$) and bright excitons ($c^\dagger_\uparrow h^\dagger_\Downarrow |s_e,s_h\rangle$ or $c^\dagger_\downarrow h^\dagger_\Uparrow |s_e,s_h\rangle$) spectral lines are sandwiched by those of trions in the two latter cases (ii-a) and (ii-b).

## C. Coulomb Interactions and Binding Energies Versus QD Shape and Size

### 1. Direct Coulomb interactions

To investigate the impact of NWQD size and shape on direct Coulomb interactions, the trade-off between two primary effects must be considered: 1) axial and lateral confinements, and 2) strain-induced potentials. Principally, enlarging the dimensions of a field-free QD leads to a reduction in the mutual interaction between two single particles, as their orbital size grows in real space. This rule of thumb, however, might not be valid for particles having a low effective mass, e.g. electron or light hole. As described in Sec. V A, electrons resist being carefully adapted to the confinement variations. For small $a_h$, electron spreads its orbital beyond the QD insertion boundaries toward the nanowire core. This is shown in Fig. 12(a) and (c) for series B-1 and B-2 NWQDs with $a_h = 0.1$, where projections of the electron $e_0$ and hole $h_0$ normalized orbitals, $\Phi_{e_0} = |\varphi_{e_0}|^2$ and $\Phi_{h_0} = |\varphi_{h_0}|^2$ are plotted along the nanowire axis.

In Fig. 12(c), positions of the $s$-shell electron and hole orbitals deviate from the NWQD center due to the piezoelectric field, producing a axially oriented dipole; see Fig 3(b) right panel. The larger portion of electron probability density $\Phi_{e_0}$ shrinks inside the NWQD when $a_h$ smoothly exceeds 0.1, yielding a denser orbital. Further enhancement of $a_h$ consistently results in the vertical spread of electron orbital, inside the QD borders this time, thus lowers the density. Further on, lateral forces induced by the surrounding shell suppress upon enlarging $a_h$. Since we have restricted our analysis to $a_h \leq 0.5$, $\Xi_{v;e,h}$ maintains dominant and determines the strength of Coulomb interactions.

With the inclusion of $V_p$ in $H_s$, $\Phi_{e_0}$ and $\Phi_{h_0}$ are not allowed to extend freely in proportion to the QD size. Piezoelectric field reduces the spatial overlap of $e_0/h_0$ orbitals $\langle s_{e_0}|s_{h_0}\rangle$, and thus the mutual attractive interaction $q_{00}^{d,eh}$. Moreover, it enhances the magnitude of repulsive interaction terms, $q_{00}^{d,ee}$ and $q_{00}^{d,hh}$, through accumulating the orbitals near the charge centers. Depending on the strength of piezoelectricity with respect to other potential terms in $H_s$, the evolution of Coulomb interaction terms versus $a_h$ is treatable as follows:

(i) The feature of being continually restricted to the QD boundaries shown by the ground state hole orbital $\Phi_{h_0}$ usually translates to a persistent reduction of $q_{00}^{d,hh}$ upon increasing the QD size. In the case of [001]-oriented NWQDs, the relaxation of confinement and growth of $V_p$ compete together. Hole orbital reacts more sensitively toward $\Xi_{v;h}$ rather than piezoelectricity, as could be concluded from Figs. 13(a) and (b), where $q_{00}^{d,hh}$ consistently drops versus the QD height. In [111]-oriented NWQDs, the relatively strong built-in piezoelectric field along the nanowire axis, however, may overcome the impact of axial relaxation for larger aspect ratios ($0.4 < a_h < 0.5$); see Fig. 13(c). We also observe that $q_{00}^{d,hh}$ follows the same trend in cone-shaped [111]-oriented NWQDs, but grows slighter for $0.4 < a_h < 0.5$ since $\Phi_{h_0}$ occupies a larger volume despite the stronger piezoelectric field; this is further detailed in (iv).

(ii) Evolution of $q_{00}^{d,ee}$ against $a_h$ could be divided in two regimes: 1) Starting from flat NWQDs, both $V_p$ and $\Xi_{v;e}$ cooperate to enhance the orbital density and therefore $q_{00}^{d,ee}$ 2) Proceeding further toward larger ratios, they begin to oppose each other, thus $q_{00}^{d,ee}$ smoothly drops.

(iii) Variation of $q_{00}^{d,eh}$ comprises both regimes of evolution discussed in (i) and (ii). Since $\Phi_{e_0}$ and $\Phi_{h_0}$ possess $D_{2d}$ and $C_{2v}$ symmetries, it seems not trivial to precisely predict how they overlap over the space. According to Fig. 13 $q_{00}^{d,eh}$ resembles $q_{00}^{d,ee}$ in terms of growing initially and receding eventually. This behavior could be inferred from Fig. 12, where the axial overlap of $\Phi_{e_0}$ and $\Phi_{h_0}$ is greater for $a_h = 0.4$ and expected to decrease for larger ratios. The average measure of electron-hole Coulomb attraction term agrees with the previous theoretically and experimentally estimated values [69] [35].

(iv) A comparison between the interaction terms in series B-1 and B-2 (C-1 and C-2) NWQDs for each specific $a_h$ clarifies the role of piezoelectricity. Larger $V_p$ existing in [001]-oriented cone-shaped NWQDs lightly increases the repulsive interactions. The QD steep side walls also make the effective volume felt by the orbitals smaller. For large $a_h$, this amplifies the effect of piezoelectricity in terms of repulsive interactions, but competitively attempts to enhance the electron-hole overlap. In [111]-oriented NWQDs, instead, the axial piezoelectric field attempts to collect $\Phi_{h_0}$ near the top interface; see Fig. 3(a) right panel. Steep sidewalls in the cone-shaped type push $\Phi_{h_0}$ toward the QD center; see Fig. 6(b). Hence, the net impact of $\Xi_{v;h}$ and piezoelectric field appears as a small suppression in both repulsive and attractive interactions compared to the cylindrical type.

(v) All direct Coulomb matrix elements decay hyperbolically ($\propto 1/D_D$) by relaxing the lateral confinement ($\Xi_{l;e,h} = \Delta_{CB,VB} + \langle H_s \rangle_{l;e,h}$) irrespective of the particle type; see Figs. 13(f) and (g). Their analogous evolution stems from the fact that effective masses of electron and hole are comparable perpendicular to the nanowire axis. The decaying behavior seems to be moderated in [100]-elongated NWQDs after a certain amount of 1D elongation (here, ~1.25 for series F). This can be explained through $e/h$ dispersions: when QD is laterally extended along [100] direction, the corresponding hole effective mass in this orientation $m_{h,[100]}^*$ becomes lighter, while $m_{h,[010]}^*$ undergoes a negligible change. The relatively steeper dispersion resulted from $m_{h,[100]}^*$ allows only dilute parts of $\Phi_{h_0}$ to extend along [100] direction, otherwise it would cost a considerable exchange of kinetic energy (see Sec. VI C 2). $\Phi_{e_0}$, on the other hand, splits parallel to the elongation axis and gets dilute along the main quantization axis [001]. The electron-induced mean-field potential felt by $e_0$ $\langle \psi_{0;\uparrow,\downarrow}^e | V_{\uparrow,\downarrow}^e | \psi_{0;\uparrow,\downarrow}^e \rangle$ or $h_0$ $\langle \psi_{0;\Uparrow,\Downarrow}^h | V_{\uparrow,\downarrow}^e | \psi_{0;\Uparrow,\Downarrow}^h \rangle$, however, turns out not to change significantly under this orbital splitting.

### 2. Dependence of electron and hole kinetic energies on orbital extent and QD size

Any effort expended by the few-particle complexes for minimizing their total energy in response to the size or shape variations is reflected in their corresponding orbital spreadout. In this regard, the amount of kinetic energy obtained by each one of the constituting particles can be estimated through its expectation value: $E_k = \langle \psi(\mathbf{k}) | H_k | \psi(\mathbf{k}) \rangle \cong \langle \psi(\mathbf{k}) | E(\mathbf{k}) | \psi(\mathbf{k}) \rangle$. $E_k$ is then directly linked to the dispersion relation $E(\mathbf{k})$ and the orbital extent in $\mathbf{k}$-space. According to this relation, sensitivity of each individual envelope function to any kind of structural variation relies not only on its effective mass, dictated by $E(\mathbf{k})$, but also upon its original size of probability density within the real space. Above relation also implies that a for a fixed value of effective mass, a larger extent of an orbital in the real space leads to a lower gain of kinetic energy.

By scrutinizing the evolution of direct Coulomb interactions in Fig. 13, we observe that $q_{00}^{d,hh}$ exhibits a greater sensitivity to the size variations for smaller $a_h$. This behavior can be explained by taking the roles of both effective mass and piezoelectricity into account: (i) having $a_h$ increased, $\Phi_{h_0}$ grows and becomes more responsive to any additional size variation. (ii) Meanwhile, the effective mass along the nanowire axis ($m_{h;[001]}^*$ or $m_{h;[111]}^*$) gets lighter due to the relaxation of $\Xi_{v;h}$ (LH percentage becomes larger). This moderates the orbital sensitivity by enhancing the potentially-available amount of $E_k$ that may be gained by any tiny spreading out. (iii) As mentioned before, the piezoelectric field attempts to squeeze $\Phi_{h_0}$ at large $a_h$, thus partially compensating the role of QD size. Eventually, having one electron and one hole involved in each mutual attractive interaction, it evolves as a mixture of electron-electron and hole-hole repulsive interactions to some extent.

To examine how QD flatness affects $E_k$ acquired by each type of carrier, we defined a renormalized orbital-dependent kinetic energy as $\xi_{h_0,e_0} = \langle H_k \rangle_{h_0,e_0} - E_{VB,CB}^0$ for $e_0$ and $h_0$ discarding the heterostructure band offsets, and calculated it for $a_h = 0.1$ to $0.5$; see Table IV. The orbital-dependent parts of electron and hole kinetic energies grow as we keep flattening the NWQDs since the associated $\rho_{e_0}(\mathbf{k})$ and $\rho_{h_0}(\mathbf{k})$ swell out. $\xi_{h_0}$ also grows given larger aspect ratios ($a_h > 0.4$) because of the further LH character the hole particle obtains. We further detail this behavior in Sec. VI C 3 where its effect on the correlation energies is discussed.

### 3. Correlation and Binding Energies

The correlation energies each ground state shares with higher excited states are essentially determined by a number of parameters including 1) size and position of single particle orbitals $\Phi_{e,h}$, 2) energy spectrum, 3) number of particles existing in the complex, and 4) kinetic energy variations $\delta E_k$. In the few-particle picture, once a carrier resting in one particular state correlates its wavefunction with other orbitals, it is looking for any deformation stabilizing the complex. This mechanism involves direct interactions established between the carrier and other single particles, i.e. how the associated orbital reshapes to minimize the total energy affected by attractive and repulsive Coulomb interactions. Returning to Fig. 13, repulsive terms, $q_{00}^{d,hh}$ and $q_{00}^{d,ee}$, are much stronger than $q_{00}^{d,eh}$ in flat NWQDs although $e_0$ and $h_0$ share rather the same centers of mass. This is due to the tight $\Xi_{v;e,h}$ which keeps the orbital sizes small. We observe larger correlation energies in this range almost for all series of NWQDs; see Fig. 14(a) to (e). According to the reasons explained in Sec. VI C 1, the interplay between the dominant $\Xi_{v;e,h}$ and piezoelectric effect causes the Coulomb interactions to approach each other while increasing $a_h$ up to 0.5. The correlation energies thus choose a descending evolution in accordance with the differences between $q_{00}^{d,hh}$, $q_{00}^{d,ee}$ and $q_{00}^{d,eh}$.

The orbital-dependent part of kinetic energy $\xi_{e,h}$ is plotted for Series B and C NWQDs in Fig. 15 based on the data given in Table IV. The reduction rate drops for both $\xi_{e_0}$ and $\xi_{h_0}$ when the size of QD increases. For $e_0$, the dispersion relation $E_{e_0}(\mathbf{k})$ is barely influenced when the orbital $\Phi_{e_0}$ slowly grows. For the hole particle, we face two competing effects: enlarging the QD axially enhances the LH character of the $h_0$ thus its dispersion $E_{h_0}(\mathbf{k})$ gets steeper. On the other hand, orbital is spreading out rapidly. The former attempts to increase $\xi_{h_0}$ while the latter acts oppositely. For $a_h < 0.4$, dispersion curve is not steep enough to overcome the effect of orbital growth, respecting the fact that hole particle accelerates gaining LH character for larger $a_h$; see Fig. 11(a). Finally for $a_h > 0.4$, kinetic energy gained by $|h_0\rangle$ orbital $\xi_{h_0}$ is enhanced due to the steeper dispersion curve.

Regardless of two above regimes of variation for holes in [001]-oriented NWQDs, they lose (or gain) less kinetic energy than electrons in response to any size variance, comparing Figs. 15(a) and (b), thus become more flexible to correlate their orbitals with excited states. Hence we generally expect $X_0^+$ (having contribution from two $h_0$ and one $e_0$) to exhibit larger correlation energy than $X_0^-$ (having contribution from two $e_0$ and one $h_0$). In contrast, hole obtains comparable kinetic energy to electron in [111]-oriented NWQDs for this range of $a_h$; see Figs. 15(c) and (d). This can be attributed to the existing axial piezoelectric field which squeezes the orbital in the real space. Consequently, the trade-off between dispersion and orbital extent is further motivated by piezoelectricity in these types of NWQDs; this means that $q_{X^-}^{corr}$ may even exceed $q_{X^+}^{corr}$ wherever the impact of orbital size becomes dominant.

Fluctuations of the hole's orbital-dependent kinetic energy $\Delta\xi_{h_0}$ for any given $a_h$ seem to become more dependent on the orbital size rather than the hole dispersion as $q_{X^+}^{corr}$ and $q_{XX}^{corr}$ start increasing for larger aspect ratios ($a_h > 0.4$). We predict they continue adding up although the total amount of correlation would be lower than the hypothetical situation when the effect of dispersion was switched off. Any lateral elongation may also increase the correlation energy for a certain complex, as it enhances the orbital size while reduces the average spacings between CB and VB energy eigenvalues. Suppression of splitting between the QD eigen-energies leads to higher density of states and thus to more available states to correlate with. The amount of correlation energy that $X_0$ collects is always smaller than the other complexes because of the lower number of carriers it contains. Conversely, $XX_0$ usually exhibits the largest correlation energy.

Binding energy is primarily determined by direct Coulomb interaction terms since they are much stronger than correlation energies in our studied NWQDs. As shown in Fig. 14, $XX_0$ and $X_0^+$ are deeply anti-binding for flat NWQDs due to their large $q_{00}^{d,hh}$. This behavior has been experimentally observed in Ref. [74]. Our range of dimensions offers no binding biexciton except for cylindrical [001]-oriented NWQDs and exclusively $a_h$ larger than ~0.45. One may reach binding regime for the other types as long as QD size, either laterally or axially, becomes larger so that direct Coulomb terms approach each other and further correlation energies are obtained. The binding energy of negative trion $X_0^-$ however behaves much more smoothly since $q_{00}^{d,eh}$ and $q_{00}^{d,ee}$ merge closely. Finally, we should notice that the exact measure of binding energy for each complex is quite fragile incorporating the realistic amount of correlation energy; however, the trend under which it changes for a homogeneous QD is explained above.

### 4. Exchange energies and spin flip time

Even though exchange interaction terms play a substantial role in the QD operation as an entangled photon source in the regular scheme, they are less important in the time reordering scheme as no fundamental restriction applies here to the energy ordering of bright excitons: exchange energies weakly contribute to the binding energies of complexes. Regarding the anisotropic exchange interaction, our calculations confirm the very small expected FSS (< 0.6 $\mu$eV within our numerical accuracy) for series B-1, C-1, C-2 and D-1 NWQDs which benefit from the elevated $D_{2d}$ and $C_{3v}$ symmetry characters [75]. Dark-bright energy splitting $\Delta_{bd}$ ranges between ~150 to 280 $\mu$eV for series B and ~270 to 410 $\mu$eV for series C, generally larger for cone-shaped NWQDs. In addition, the Rashba spin-orbit interaction originated from the anisotropy in series C NWQDs offers an extra source of spin dephasing increasing the relaxation rates [76].

In order to quantitatively demonstrate the impact of quantization and symmetry on spin relaxation times, we examined series C-1 and C-2 NWQDs. Let us consider the first mechanism in Sec. III. B. In series C-1 both $\delta_s$ (FSS) and $\Delta_{bd}$ increase with respect to $a_h$, leading to an abrupt drop in $\tau_{1,a}$ as it changes proportional to $1/\Delta_{bd}^2$ and $1/\delta_s^3$; see Fig. 16(a) inset. In series C-2, $\delta_s$ is sub-$\mu$eV considered equal to the upper limit of 0.6 $\mu$eV in Fig. 16, and only $\Delta_{bd}$ grows with $a_h$, thus the sensitivity of $\tau_{1,a}$ against $h_D$ drops considerably; see Fig. 16(b). Regardless of QD type, $\tau_{1,a} \gg \tau_X$ indicating that simultaneous spin flip processes ($|s_{e,\uparrow},s_{h,\Downarrow}\rangle \leftrightarrow |s_{e,\downarrow},s_{h,\Uparrow}\rangle$) rarely randomize the phase difference between the two bright exciton decay paths.

Second mechanism primarily relies on lateral quantization (or equivalently $D_D$) along with $\Delta_{bd}$ rather than FSS, as the transition occurs between bright $X_B$ and dark excitons $X_D$. Each bright exciton $|s_{e,\uparrow},s_{h,\Downarrow}\rangle$ ($J=-1$) or $|s_{e,\downarrow},s_{h,\Uparrow}\rangle$ ($J=1$) can switch to either $|s_{e,\uparrow},s_{h,\Uparrow}\rangle$ ($J=2$) or $|s_{e,\downarrow},s_{h,\Downarrow}\rangle$ ($J=-2$) dark states only by flipping one electron or hole spin. Figs. 16(c) and (d) show $\tau^e_{J=1\to 2}$ and $\tau^h_{J=1\to -2}$ for $X_B \to X_D$ transition in series C-1 and C-2 NWQDs. Two QD diameters are examined, 20 and 30 nm, while the height $h_D$ is kept fixed equal to 4 nm. The corresponding $X_B$-$X_D$ exchange splittings are $\Delta_{bd} = 242$ and 93 $\mu$eV calculated for series C-1 and $\Delta_{bd} = 345$ and 136 $\mu$eV for series C-2 NWQDs. Results show that spin flip process involving exciton-bound electron generally occurs quicker than the one of the hole. Also, QD having larger diameter exhibits longer $\tau_{1,b}$ when its height is fixed. This is interpretable when $\hbar\omega_0 \gg \Delta_{bd}$ and thus $\tau_{1,b} \propto D_D^8/\Delta_{bd}^5$, implying that the strength of axial quantization enhances the spin coherence. $\Delta_{bd}$ is sufficiently small in series C-1 NWQDs so that $\tau_{1,b} \gg \tau_X$ even for higher temperatures. The larger $X_B$-$X_D$ splitting in series C-2 NWQDs, however, decrease spin flip times significantly where $\tau_{1,b}$ becomes comparable to the typical exciton lifetimes; see $\tau^e_{J=1\to 2}$ (solid line) in Fig. 16(d) where $a_h = 0.2$ ($h_D = 4$ nm and $D_D = 20$ nm). Spin flip times $\tau^{e,h}_{J=2\to 1}$ of $X_D \to X_B$ transition present similar trend in terms of $\tau^h_{J=2\to 1} > \tau^e_{J=2\to 1}$ and $\tau_{1,b}$ being in the order of $\tau_X$ for series C-2 NWQDs. As suggested by Figs. 16(e) and (f), exciton dark state $X_D$ remains resistant to spin flip in very low temperatures since the process is phonon-mediated

### D. Controlling Coulomb Interactions via Electric Field in NWQDs

The binding energies of few-particle complexes in our studied NWQDs are strongly governed by direct Coulomb interaction terms rather than correlation energies. Owing to the larger correlation energy of biexciton $XX_0$ compared to bright exciton $X_{0,B}$, we may find its energy level appearing below that of $X_{0,B}$ when QD size brings the amount of direct interactions to very similar values. A controllable external perturbation, such as Stark effect, then enables us to delicately tailor these terms and, therefore, the resultant binding energy $\delta(nX^m)$. In general, exerting an external electric field introduces the following effects: 1) It redshifts and blueshifts the $e_0$ and $h_0$ energies, respectively, thus attempting to reduce the s-shell transition energy $E_{e_0 \to h_0}$ from the single particle point of view. 2) Spatially separates the electron and hole orbitals, hence, diminishes their attractive interaction $q^{d,eh}$. This reduction in the exciton binding energy, partially cancels the redshift in $E_{e_0 \to h_0}$. 3) It can be visualized as an external piezoelectric field, capable of enhancing the repulsive interaction terms by squeezing $\Phi_e$ and $\Phi_h$. 4) Renormalizes the binding energies by manipulating $\Phi_e$ and $\Phi_h$ along with the mean-field potentials. 5) A lateral electric field ruins the symmetry character of wavefunctions. Therefore, under a sufficiently high local electric field, FSS becomes large enough to exclude the possibility of employing the regular scheme. Moreover, a lateral electric field can decrease the radiative recombination lifetime of ground state complexes, yielding the emergence of higher-order complexes in which one or a number of single particles reside p-states [8].

In this section we study two specific situations where perpendicular (lateral) and parallel (axial) electric fields are applied to [001]- and [111]-oriented NWQDs possessing small $XX_0$ binding energies. According to the earlier discussions in Sec. VI C 3, biexciton is 2.84 meV binding in series B-1 NWQD with $a_h = 0.5$; see Fig 15(a). Though a part of this energy comes from correlation terms, zero $XX_0$ binding energy could be realized through spatially splitting $e_0$ and $h_0$, and henceforth delicately reducing $q^{d,eh}_{00}$. Introducing external perturbation $\langle \psi_{e,h}|eV_{ext}|\psi_{e,h}\rangle$ into the total hamiltonian $H_k + H_s$ for this purpose requires a relatively strong electric field due to the considerable difference between dielectric constants inside ($\varepsilon_{in}$) and outside ($\varepsilon_{vac}$) the nanowire. Moreover, mean field potentials produced by each particle are large scale, thus exhibit trivial deformations in response to the electric field-induced orbital distortions.

We modeled an artificial lateral gate by introducing a constant [100]-directed electric field, labeled as $E_x$ in Fig. 17, into the structure and recalculated the single particle orbitals for the average internal electric fields $\langle E_{[001]}\rangle_{V_D}$ equal to 3.61, 10.8 and 18 kV/cm ($V_D$ is the QD volume). These values correspond to 25, 75 and 125 kV/cm external fields, respectively. By performing CI calculations we found that along with separating $e_0$-$h_0$ orbitals, Stark effect also enhances the repulsive interactions, $q^{d,hh}_{00}$ and $q^{d,ee}_{00}$, by shrinking $\Phi_{e_0}$ and $\Phi_{h_0}$, thus binding energies quickly approach zero during the dipole formation. Expectedly, $q^{d,ee}_{00}$ and $\delta(X_0^-)$ have the least variations. Correlation energies very slightly decrease in response to the $\langle eV_{ext}\rangle \approx 36$ meV polarizing potential dropping across the QD ($V_{ext} \approx E_{[100]} D_D$) because 1) orbitals are tightened and 2) LH character is enhanced in the presence of strong electric field as shown in Table V. Moreover, anisotropic long range exchange interaction $\langle e_{0;\uparrow}h_{0;\Downarrow}|C|e_{0;\downarrow}h_{0;\Uparrow}\rangle$ grows rapidly as a result of symmetry breaking and reorders the fine structure energy $\delta_s$.

A parallel electric field is also able to restore the color coincidence between $XX_0$ and $X_0$ without breaking the orbital symmetries. Depending on the respective positions of single particle orbitals, it can shift $XX_0$ resonance into or bring it out of the binding regime. In the particular case of [111]-oriented NWQDs, an axial dipole already exists as a result of piezoelectric potential (see Fig. 3), pushing $XX_0$ resonance toward antibinding region: $q_{00}^{d,hh} + q_{00}^{d,ee} > 2q_{00}^{d,eh} + \Delta q_{X_0,XX_0}^{corr}$. An equivalent parallel electric field, $|E_{ext}| \approx E_{[111]}^{piezo}$, but oppositely directed then may compensate the built-in polarization. $|e_0\rangle$ and $|h_0\rangle$ orbital distributions versus $E_{ext}$ are demonstrated in Fig. 18(a) for series B-2 NWQD with $a_h = 0.5$ where $\delta(XX_0) = -3.37$ meV. $|h_0\rangle$ orbital touches the top interface in relatively weak electric fields. By increasing $E_{ext}$ up to 17.4 kV/cm, axial component of the internal piezoelectric field is almost canceled out and $\Phi_{h_0}$ occupies the entire QD space encompassing $\Phi_{e_0}$. A temporary rise in the spatial overlap and $q_{00}^{d,eh}$ then occurs, leading to a binding $XX_0$ [$\delta(XX_0) \approx 0.8$ meV]; see Fig. 18(c). Note that $q_{00}^{d,hh}$ experiences a transient drop until electric field becomes strong enough, $E_{ext} = 26.2$ kV/cm, to collect $\Phi_{h_0}$ near the bottom interface. At the same time we observe a trivial minimum happening in $\xi_{h_0}$ once $\Phi_{h_0}$ spreads out axially. In contrast, $\Phi_{e_0}$ and thus its resultant mean-field potential $V_{0,\sigma}^e$ stay stiff, while the associated kinetic energy $\xi_{e_0}$ delivers very limited variations (< 0.1 meV).

## VII. Summary

Electronic and spectroscopic characteristics of experimentally realizable NWQDs are investigated through establishing a connection between their morphological parameters and physical observables. With the aid of CI calculations we showed that sub-$\mu$eV anisotropic exchange interaction makes NWQDs potential candidates as entangled photon source. Two level of analysis have been performed: first, single particle level in which symmetry characters of s- and p-shell orbitals, overlap of e/h probability densities and their spreadout in k-space are discussed in detail. We quantitatively demonstrated how strain-induced potentials (including $\langle V_{Hy}\rangle_{e,h}$, $\langle V_{Bi}\rangle_{e,h}$ and $\langle V_p\rangle_{e,h}$) change the orbital symmetry, e/h dipole and s-shell hole character. It is particularly shown that flat NWQDs, even if suffering from axial anisotropy, gain low contribution from LH and SO characters in their s-shell oscillator strength.

Second, few-particle level, where the energy separation of s-shell excitonic resonances $X_0$, $XX_0$ and $X_0^\pm$ were elaborated in terms of direct, exchange and correlation terms. We defined and calculated orbital dependent kinetic energy $\langle H_k\rangle_{e_0,h_0} - E_{CB,VB}^0$ to explain orbital deformations and track interaction terms in the few particle complexes: a) the energy difference between attractive ($q^{eh}$) and repulsive ($q^{ee}$ and $q^{hh}$) interaction term drop when axial and lateral localizations become comparable. b) mean field potentials barely obey tiny structural details. c) correlation energies get important when axial localization is considerably dominant. $XX_0$ and $X_0^+$ correlation energies also grow when axial confinement is sufficiently relaxed. d) accumulating correlation energy is prohibited wherever orbital distortion causes non-negligible gain of kinetic energy. We specifically examined the variations of $XX_0$ and $X_0^\pm$ binding energies versus the vertical aspect ratio: achieving binding $XX_0$ and $X_0^\pm$ is unlikely in flat NWQDs where $q^{hh}$ noticeably exceeds $q^{ee}$ and $q^{eh}$. For the same reason, time reordering scheme cannot be implemented with flat NWQDs.

Spin relaxation times were calculated according to fine structure splittings of excitonic resonances, $\Delta_{bd}$ and $\delta_s$. We found that the spin flip process which may destroy the phase of exciton-photon wavefunction during $XX_0$ recombination is dependent on $\Delta_{bd}$ rather than FSS. Also, exciton-bound hole spin exhibits more stability than electron spin.

Binding energies of $XX_0$ and $X_0^\pm$ were manipulated by applying axial and lateral electric fields. $XX_0$ binding energy could be effectively erased merely through adjusting direct interactions since correlations turned out to be almost unaffected under external electric field. We particularly showed that it would be possible to have $\delta(XX_0) = 0$ and $\delta_s = 0$ simultaneously in [111]-oriented NWQDs. This could lead to realization of an entangled photon source emitting photons with an identical color in every sequence.

### Acknowledgements

This research is supported by NSERC-NRC-BDC and IQC funding. We thank Jonathan Baugh form the Coherent Spintronics group at IQC for his fruitful discussions. We also thank Christopher Haapamaki and Ray LaPierre from CEDT at McMaster University for sharing their experience on growing NWQDs. M. Khoshnegar also thanks Irene Goldthorpe for her comments on the practical issues of shell passivation. We acknowledge IST services at University of Waterloo for supporting computational facilities.

## APPENDIX A: MODELING AND NUMERICAL DETAILS

In the following, we describe the different methods of calculation used in this paper.

(i) Strain calculation: to solve for the strain distribution $\varepsilon_{ij}(x,y,z)$ the strain energy is calculated and minimized for core-shell nanowires using CM method and generalized minimal residual algorithm (GMRES). The bottom surface is kept fixed (no displacement) while it is assumed that QD stress relaxes at the top surface for some specific points at shell perimeter ($\delta R_x$, $\delta R_y$ and $\delta R_z = 0$). Nonuniform discretization has been utilized for better convergence: refined meshing has been applied to QD and shell regions.

(ii) Piezoelectric potential $V_p$: in order to solve the Poisson's equation associated with the piezoelectric charge density $\nabla \cdot (\boldsymbol{P}_a + \boldsymbol{P}_b)$, we utilized the finite element method on an adapted nonuniform three-dimensional grid with gradual growth of the element length outside the shell. Considering a $150 \times 150 \times 150$ nm$^3$ cell, potential was computed on, by average, $2.5 \times 10^7$ elements including $1.26 \times 10^7$ elements inside and $1.24 \times 10^7$ elements outside the nanowire heterostructure (number of elements slightly changes in different geometries). Periodic boundary conditions were applied along the nanowire axis while Dirichlet boundary conditions were assigned to the cell side walls (perpendicular to the growth direction).

(iii) Single particle states: eight-band **k.p** hamiltonian is diagonalized while applying Dirichlet boundary condition at the nanowire sidewalls (external boundaries of the shell region) and preserving the continuity of probability current across the interior interfaces. The nanowire length was specified to be 150 nm so that the QD energy levels and their envelope functions are not affected by the periodic or Dirichlet boundary conditions at the two hypothetical nanowire ends. The average length of elements inside the QD region was chosen as small as 0.4 nm. This small length was preserved inside shell and nanowire regions close to the QD, growing very slowly toward the nanowire terminating boundaries. GMRES algorithm along with incomplete LU (ILU) pre-conditioner have been employed to solve $H_{\mathbf{k.p}}$. **k.p** parameters used in this paper have been taken from Refs. [51] and [58].

(iv) Mean field potentials: to obtain mutual Coulomb interactions, the mean field potential caused by each single particle ($V_{\sigma_1(\sigma_2)=\uparrow,\downarrow,\Uparrow,\Downarrow}^{m(n)=e,h}$) shall be calculated through solving Poisson's equation:

$$-\nabla \cdot \left( \varepsilon_0 \varepsilon_r \nabla V_{\sigma_1,\sigma_2}^{m,n} \right) = e^2 \left\langle \psi_{\sigma_1}^m \middle| \psi_{\sigma_2}^n \right\rangle \quad \text{(A1)}$$

where $m$ ($n$) and $\sigma_1$ ($\sigma_2$) sweep over all possible levels and spin configurations. Boundary conditions similar to (ii) are applied here.

## APPENDIX B: CONSIDERATIONS IN CONFIGURATION INTERACTION METHOD

The strategy in CI method is to assume each excitonic wavefunction as a linear combination of correlated terms in the form of Slater determinants $|\Theta_{\alpha,\beta}\rangle$:

$$\left| \Psi^{nX^m} \right\rangle = \sum_{\substack{l,\sigma_e \\ p,\sigma_h}} \gamma_{\alpha_{l;\sigma_e},\beta_{p;\sigma_h}}^{nX^m} \left| \Theta_{\alpha_{l;\sigma_e},\beta_{p;\sigma_h}}^{nX^m} \right\rangle$$

where the expansion coefficients $\gamma_{\alpha,\beta}$ depend on the number of constituting particles in the few-particle state $nX^m$, $\alpha_l = \{e_{0;\uparrow}, e_{0;\downarrow}, e_{1;\uparrow}, e_{1;\downarrow}, ..., e_{l,\sigma}\}$ and $\beta_p = \{h_{0;\uparrow}, h_{0;\downarrow}, h_{1;\uparrow}, h_{1;\downarrow}, ..., h_{p,\sigma}\}$, and the size of basis set. The number of **k.p** orbitals contributing in $|\Theta_{\alpha,\beta}\rangle$ determines the speed of convergence and accuracy of correlation energies. Type of the basis may also include unbound states existing in the nanowire (obviously not the common **k.p** spurious states). We, however, discard these unbound states since the size of eigenvalue matrices grow rapidly with the number of bases. The weight of correlation coming from the excited shells, $|p_i\rangle$ and $|d_i\rangle$, in the expansion coefficients depends on their energy spacing together with the symmetry of corresponding orbitals. Larger energy spacing between the conduction levels then allows for truncating to lower excited states. Based on the energy splittings between the higher shells we sometimes changed the number of single particle levels involved in the calculations for different series of NWQDs, but it was always maintained above 7 and 9 ($l = 14$ and $p = 18$ states by spin inclusion) for $e/h$, respectively. After expanding the multiexciton hamiltonian of Eq. (8) on the new Hilbert space of all possible configurations $\{l, p, \sigma_e, \sigma_h\}$, few-particle resonances were calculated by numerical diagonalization.

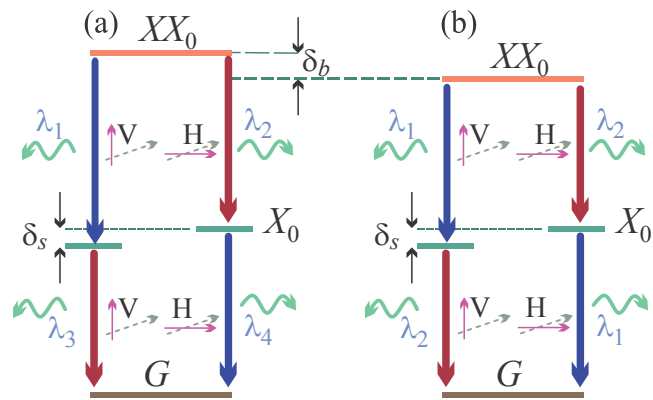

FIG. 1. QD ground state biexciton ($XX_0$)-exciton ($X_0$) cascade recombination process. Left panel: diagram of the polarized entangled photon generation in the regular scheme. Recombination diagram exhibits four distinct photon energies owing to the fine structure splitting $\delta_s$ and biexciton binding energy $\delta_b$. Right panel shows the energy diagram required for the time reordering scheme where biexciton binding energy has been removed, thus only two photon colors exist.

Single and Few-Particle States in Core-Shell NWQDs

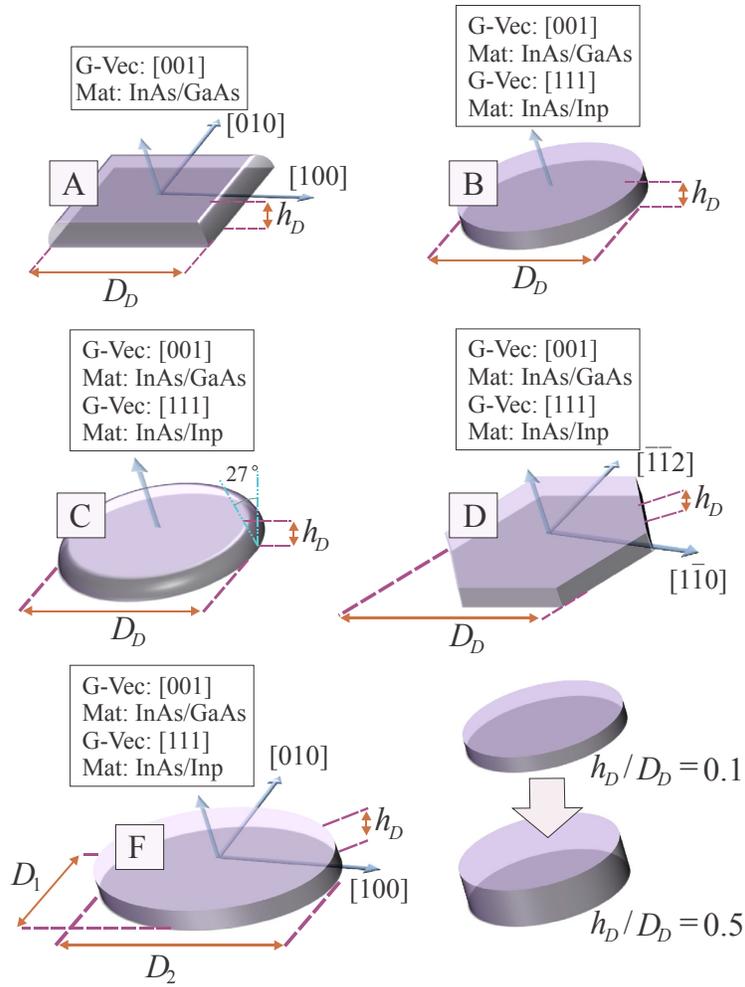

FIG. 2. Shape of the NWQD insertions modeled in this paper: A) truncated pyramidal. B) Cylindrical, C) cone-shaped, D) hexagonal and F) elongated NWQDs. Growth direction (G-Vec) and QD/nanowire materials (Mat) are specified in each case. Vertical aspect ratio varies between 0.1 and 0.5 for series B to D NWQDs. Nanowire core and shell regions are not shown here.

M. Khoshnegar et al

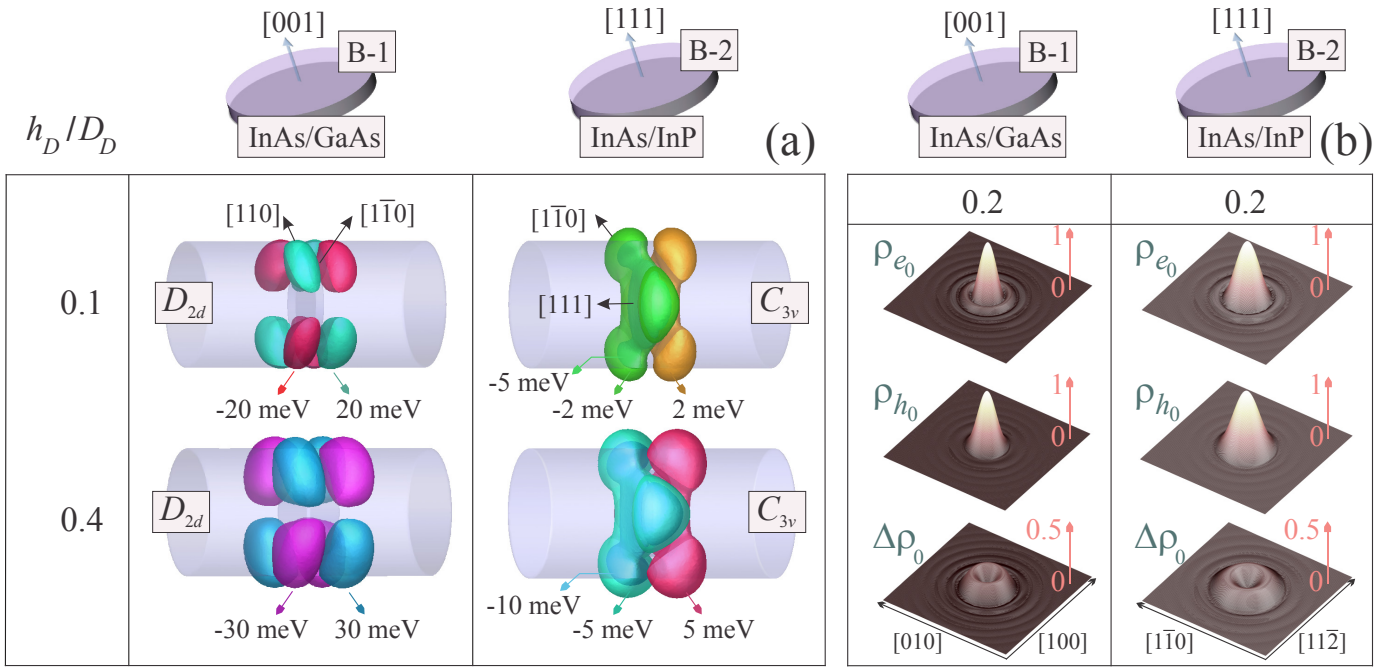

FIG. 3. (a) Piezoelectric potential in series B-1 and B-2 NWQDs illustrated for $a_h$ equal to 0.1 and 0.4. left panel: 20 (30) and −20 (−30) meV isosurfaces shown for series B-1 NWQDs. Potential has $D_{2d}$ symmetry since the top and bottom isosurfaces are equivalent $\langle V_p^U \rangle = -\langle V_p^D \rangle$ and $\langle V_p^U \rangle = e^{i\pi/2}\langle V_p^D \rangle$. Right panel: 2 (5), −2 (−5) and (−5) −10 meV isosurfaces for series B-2 NWQDs. Piezoelectric potential exhibits $C_{3v}$ symmetry; $\langle V_p^{U,D} \rangle = e^{i2\pi/3}\langle V_p^{U,D} \rangle$ and $\langle V_p^U \rangle = -e^{i\pi/3}\langle V_p^D \rangle$. (b) Electron, hole and differential charge densities shown in k-space for seris B-1 (left panel) and B-2 (right panel) with $a_h = 0.2$. Fourier transform is plotted on (001) and (111) planes for series B-1 and B-2, respectively.

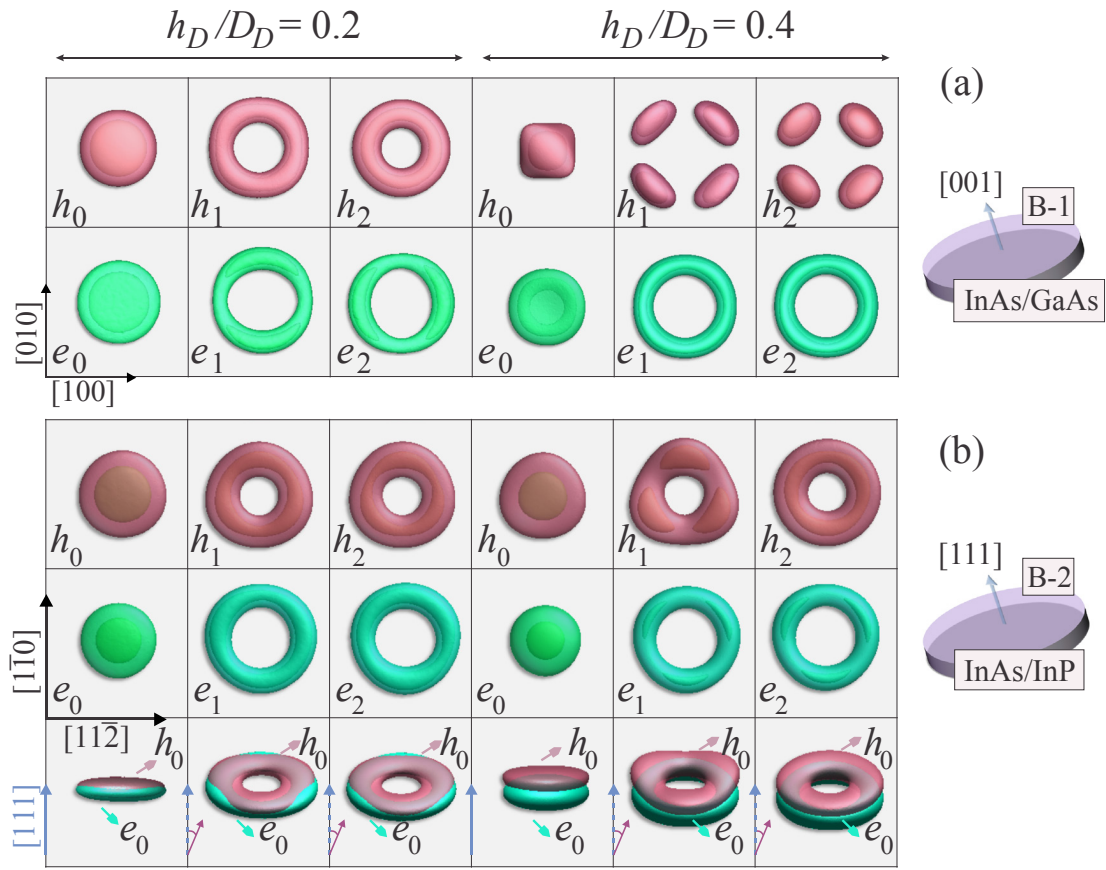

FIG. 4. Comparison of the *s*-shell and *p*-shell electron and hole orbitals in (a) series B-1 and (b) series B-2 NWQDs for $a_h$ equal to 0.2 and 0.4. (a) Series B-1 electron and hole orbitals represent $D_{2d}$ symmetry. Symmetry character becomes clearer for $a_h = 0.4$ since the tightness of $\Xi_{v;e,h}$ is reduced and orbitals reshape with less restriction. (b) Probability densities exhibit $C_{3v}$ symmetry in series B-2 NWQDs irrespective of $a_h$. Lower panel: side view shows the relative position of electron and hole orbitals. For the sake of clarity, the coordinate system is slightly rotated for *p*-states. Increasing the QD height leads to the spatial separation of electron and hole probability densities.

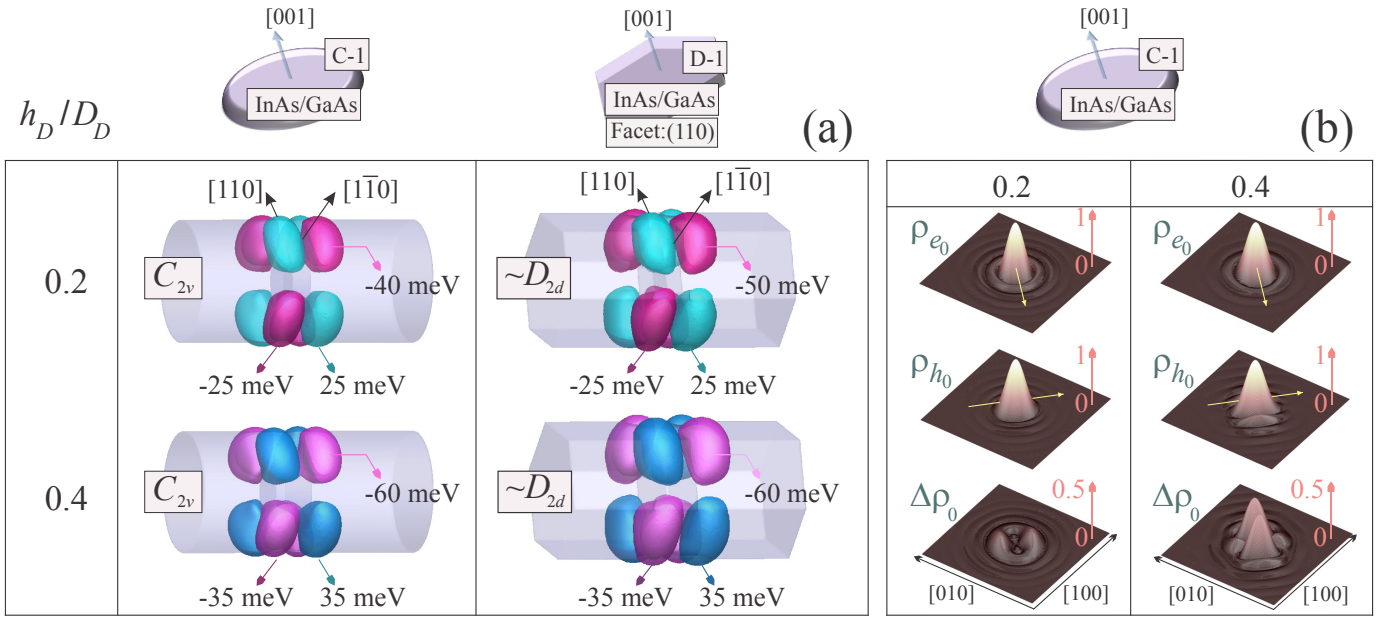

FIG. 5. (a) Piezoelectric potential in series C-1 and D-1 NWQDs illustrated for $a_h$ equal to 0.2 and 0.4. Left panel: 25 (35), −25 (−35) and (−40) −60 meV isosurfaces shown for series C-1 NWQDs. Potential has $C_{2v}$ symmetry since the amplitudes of top and bottom isosurfaces are not equivalent; $\langle V_p^U \rangle \neq -\langle V_p^D \rangle$ and $\langle V_p^U \rangle \neq e^{i\pi/2}\langle V_p^D \rangle$. Right panel: 25 (35), −25 (−35) and (−50) −60 meV isosurfaces for hexagonal NWQDs with (110) side facet. Piezoelectric potential exhibits $C_{2v}$ (~$D_{2d}$) symmetry; $\langle V_p^U \rangle = -\langle V_p^D \rangle$ but $\langle V_p^U \rangle \neq e^{i\pi/2}\langle V_p^D \rangle$. (b) Electron, hole and differential charge denisties shown on (001) plane for seris C-1 NWQDs with $a_h$ = 0.2 and 0.4. All densities are polarized due to the $C_{2v}$ symmetry of orbitals in real space.

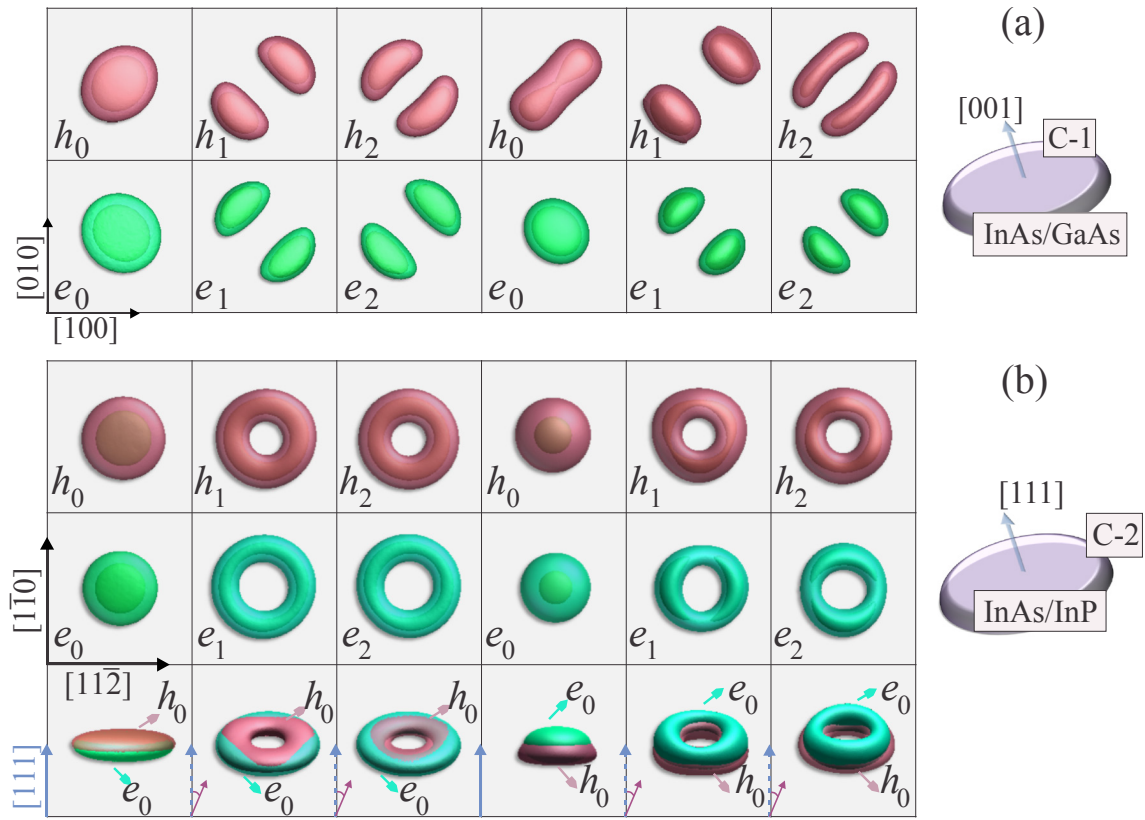

FIG. 6. *S*-shell and *p*-shell electron and hole orbitals in (a) series C-1 and (b) series C-2 NWQDs for $a_h$ equal to 0.2 and 0.4. (a) Symmetry of the single particle orbitals decrease in cone-shaped NWQDs ($C_{2v}$) due to the pronounced role of nonsymmetric piezoelectric potential. b) *S*- and *p*-shell orbitals of hole and *s*-shell orbital of electron represent $C_{3v}$ symmetry. Lower panel: probability density of electrons and holes do not have the maximum spatial overlap as a result of internal electric field.

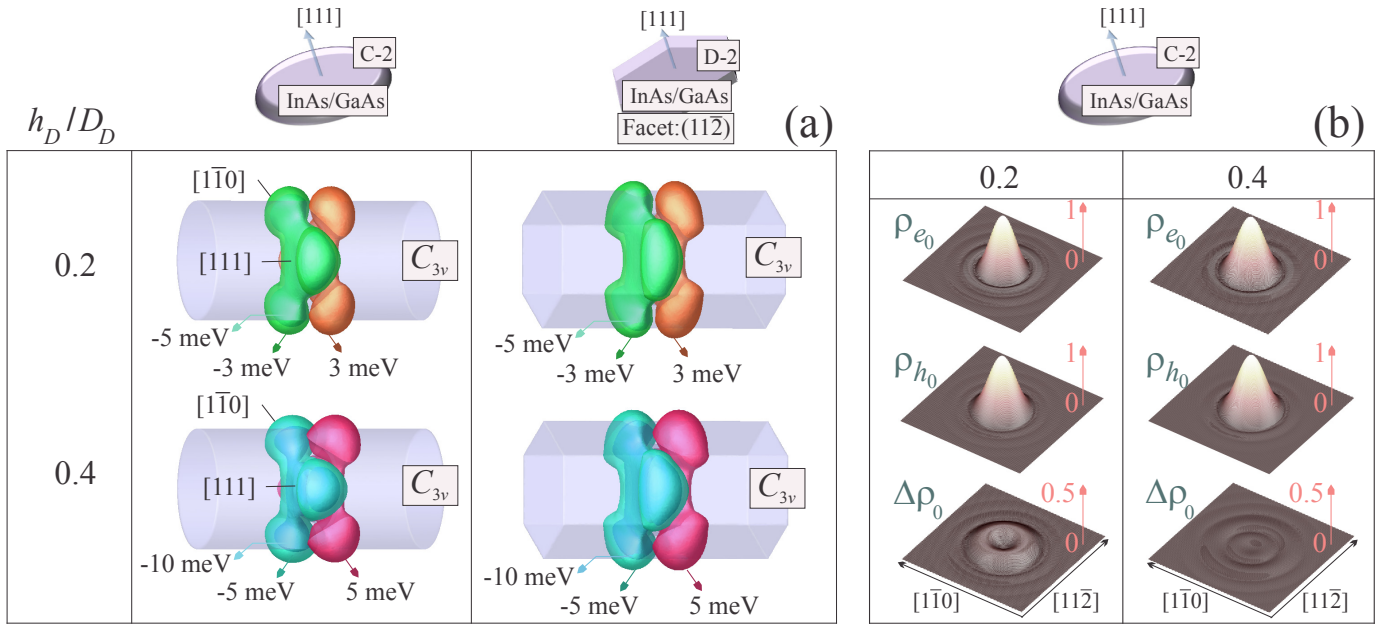

FIG. 7. (a) Piezoelectric potential in series C-2 and D-2 NWQDs illustrated for $a_h$ equal to 0.2 and 0.4. Left panel depicts −3, 3 and −5 meV isosurfaces for series C-2 NWQDs presenting $C_{3v}$ symmetry; $\langle V_p^{U,D} \rangle = e^{i2\pi/3} \langle V_p^{U,D} \rangle$ and $\langle V_p^U \rangle = -e^{i\pi/3} \langle V_p^D \rangle$. Right panel: −5, 5 and −10 meV isosurfaces for series D-2 QDs with (11$\bar{2}$) side facet. Piezoelectric potential possesses $C_{3v}$ symmetry. (b) Electron, hole and differential charge denisties shown on (111) plane for seris C-2 NWQDs with $a_h$ = 0.2 and 0.4. $\Delta\rho_0$ is suppressed for $a_h$ = 0.4 since $\Xi_{v;h}$ pushes $|h_0\rangle$ orbital against built-in piezoelectric field.

Single and Few-Particle States in Core-Shell NWQDs

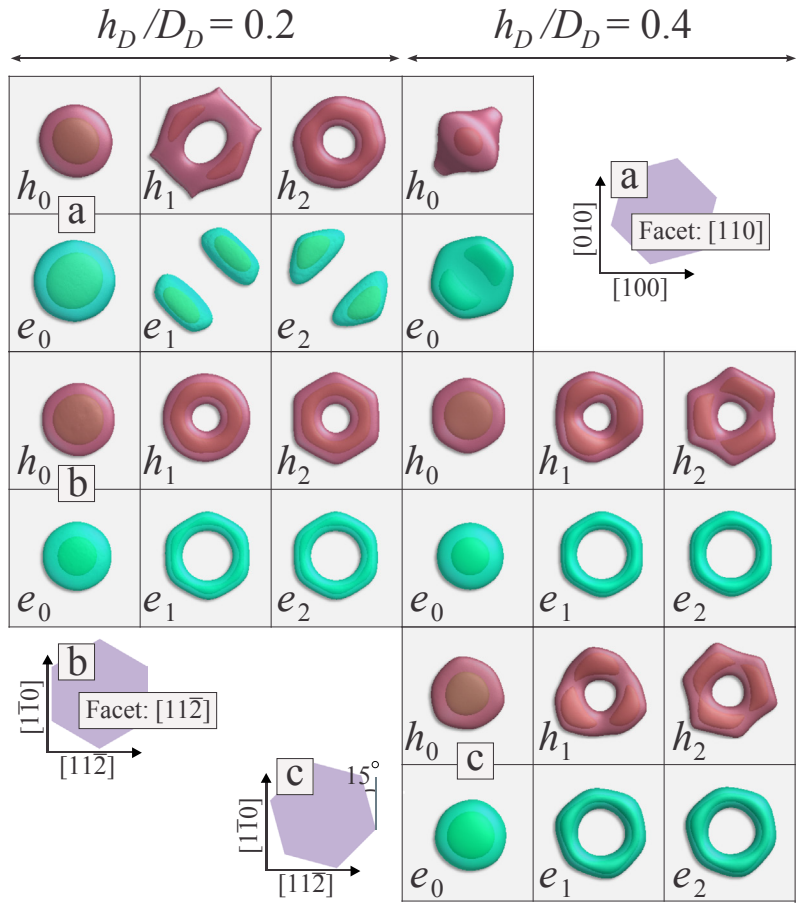

FIG. 8. Electron and hole $s$- and $p$-shells orbitals shown for $a_h = 0.2$ and 0.4 in (a) [001]- and (b), (c) [111]-oriented NWQDs with hexagonal cross section. Cross section is 15° rotated in (c) to exemplify a more general case. (a) Despite a $\sim D_{2d}$ symmetry exhibited by the piezoelectric potential, orbitals show $C_{2v}$ character especially in $p$-states. (b) and (c), both electron and hole $s$- and $p$-shell orbitals represent $C_{3v}$ symmetry. Due to the axial isotropy, the QD height is no longer a destructive factor in the global symmetry and merely adjusts the orbital density. Electron $p$-shell orbitals show almost identical spatial extent, a typical signature of small $p$-state splitting and elevated symmetry.

Table I. Expectation values of piezoelectric $V_p$, hydostatic $V_{Hy}$ and biaxial $V_{Bi}$ potentials for ground state electrons and holes in series B-1 and C-1 QDs. $\langle V_p \rangle_{e,h}$ reaches saturation in series C-1 for $a_h \geq 0.4$ due to the competition between geometry confinement and piezoelectric effect.

|  | $a_h$ | $\langle V_p \rangle_{e_0}$ (meV) | $\langle V_p \rangle_{h_0}$ (meV) | $\langle V_{Hy} \rangle_{e_0}$ (meV) | $\langle V_{Hy} \rangle_{h_0}$ (meV) | $\langle V_{Bi} \rangle_{h_0}$ (meV) |
|---|---|---|---|---|---|---|
| Series B-1 | 0.1 | −0.09 | 0.12 | 295 | −34.9 | 148 |
|  | 0.2 | −0.33 | 0.60 | 325 | −34.1 | 136 |
|  | 0.3 | −0.55 | 1.77 | 301 | −29.6 | 112 |
|  | 0.4 | −0.74 | 4.37 | 267 | −24.5 | 94.2 |
|  | 0.5 | −0.93 | 9.26 | 234 | −20.6 | 83.2 |
| Series C-1 | 0.1 | −0.10 | 0.15 | 296 | −34.6 | 147 |
|  | 0.2 | −0.33 | 0.83 | 329 | −33.5 | 135 |
|  | 0.3 | −0.76 | 3.37 | 308 | −28.8 | 115 |
|  | 0.4 | −0.71 | 6.01 | 276 | −25.3 | 105 |
|  | 0.5 | −0.68 | 6.56 | 241 | −23.6 | 103 |

Table II. Expectation values of piezoelectric and hydostatic potentials for ground state electrons and holes in series B-2, C-2 and D-2 QDs. Expectation value of biaxial strain is trivial diagonal components of strain tensor become equal in weight when growth direction is [111]. Sign of $\langle V_p \rangle_{e,h}$ is inverted in series C-2 since axial confinement pushes probability denisties angaints piezoelectric field.

|  | $a_h$ | $\langle V_p \rangle_{e_0}$ (meV) | $\langle V_p \rangle_{h_0}$ (meV) | $\langle V_{Hy} \rangle_{e_0}$ (meV) | $\langle V_{Hy} \rangle_{h_0}$ (meV) |
|---|---|---|---|---|---|
| Series B-2 | 0.1 | −0.24 | 0.41 | 136 | −21.4 |
|  | 0.2 | −0.52 | 1.48 | 157 | −18.7 |
|  | 0.3 | −0.58 | 2.62 | 147 | −15.7 |
|  | 0.4 | −0.46 | 3.77 | 131 | −13.3 |
|  | 0.5 | −0.28 | 5.11 | 115 | −11.9 |
| Series C-2 | 0.1 | −0.07 | 0.54 | 139 | −21.1 |
|  | 0.2 | 0.18 | 1.6 | 155 | −18.2 |
|  | 0.3 | 0.78 | 1.73 | 146 | −15.1 |
|  | 0.4 | 1.46 | −0.03 | 131 | −12.5 |
|  | 0.5 | 2.24 | −1.84 | 116 | −10.9 |
| Series D-2 | 0.1 | −0.24 | 0.44 | 137 | −21.6 |
|  | 0.2 | −0.54 | 1.61 | 160 | −19.0 |
|  | 0.3 | −0.61 | 2.89 | 150 | −16.1 |
|  | 0.4 | −0.48 | 4.15 | 134 | −13.6 |
|  | 0.5 | −0.27 | 5.49 | 118 | −10.9 |

Single and Few-Particle States in Core-Shell NWQDs

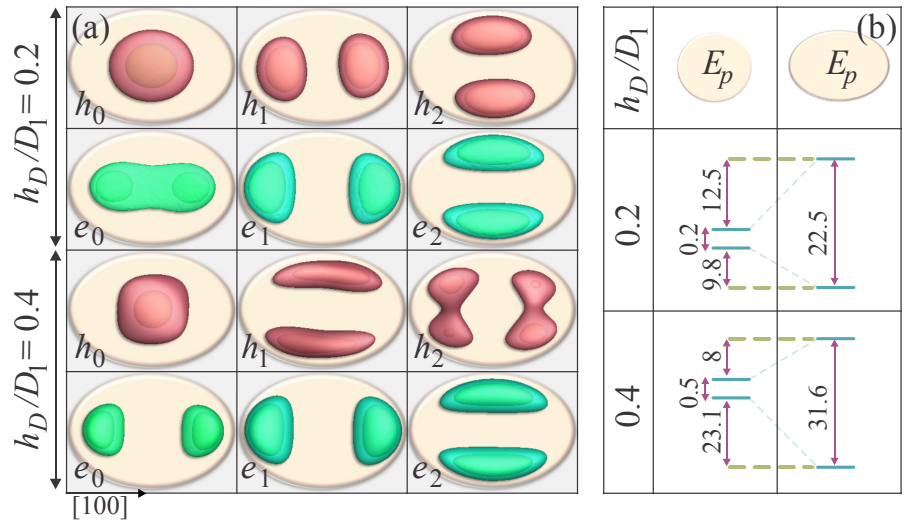

FIG. 9. (a) $S$- and $p$-shell electron and hole orbitals in a [100]-elongated (50%) NWQD grown along [001] orientation; $D_1$ = 20 nm and $D_2$ = 30 nm. Geometry dictates $C_{2v}$ character for all orbitals irrespective of the strain-induced potentials. Electron orbitals are rather insensitive to the QD height as they resist piezoelectric potential variations. (b) Electron $p$-state splitting as a function of $a_h$. Values are given in meV.

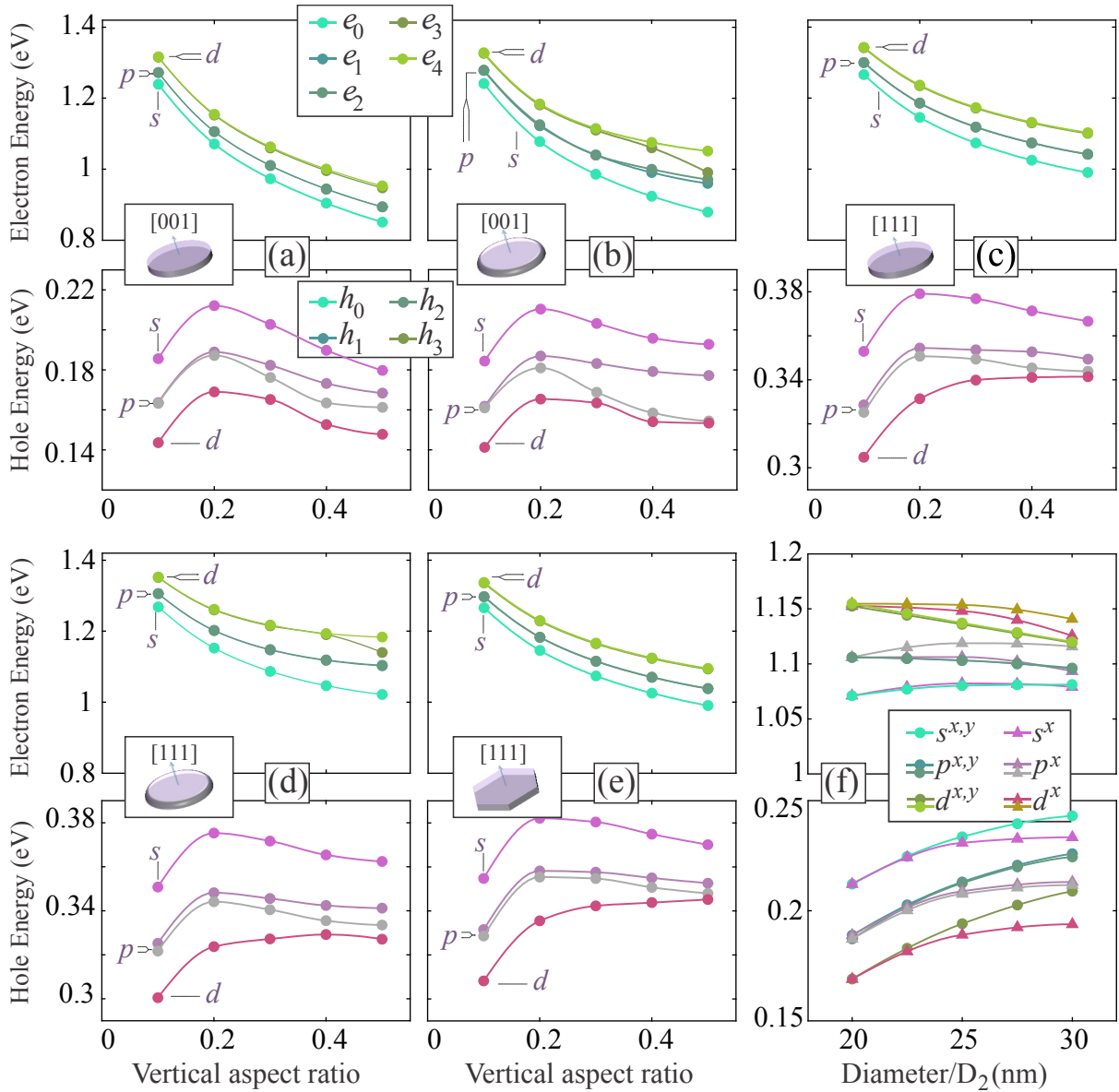

FIG. 10. Electron and hole energies at the single particle level shown for series (a) B-1, (b) B-2, (c) C-1, (d) C-2, (e) D-2 QDs against vertical aspect ratio. Reference energies are the valence band edges of homogeneous GaAs and InP nanowires; i.e. $E^0_{VB,\text{GaAs}}$ and $E^0_{VB,\text{InP}}$. Curves are interpolated data displaying the evolution of single particle energies. (f) Effect of lateral enlargement for series B-1 (F) NWQDs: $s$, $p$ and $d$-shell energies are illustrated versus the QD diameter ($D_2$) for uniform ($x$-directed) elongation. "$x$" and (or) "$y$" state that NWQD is elongated in [100] and (or) [010] crystallographic directions. The QD height is kept fixed equal to 4 nm.

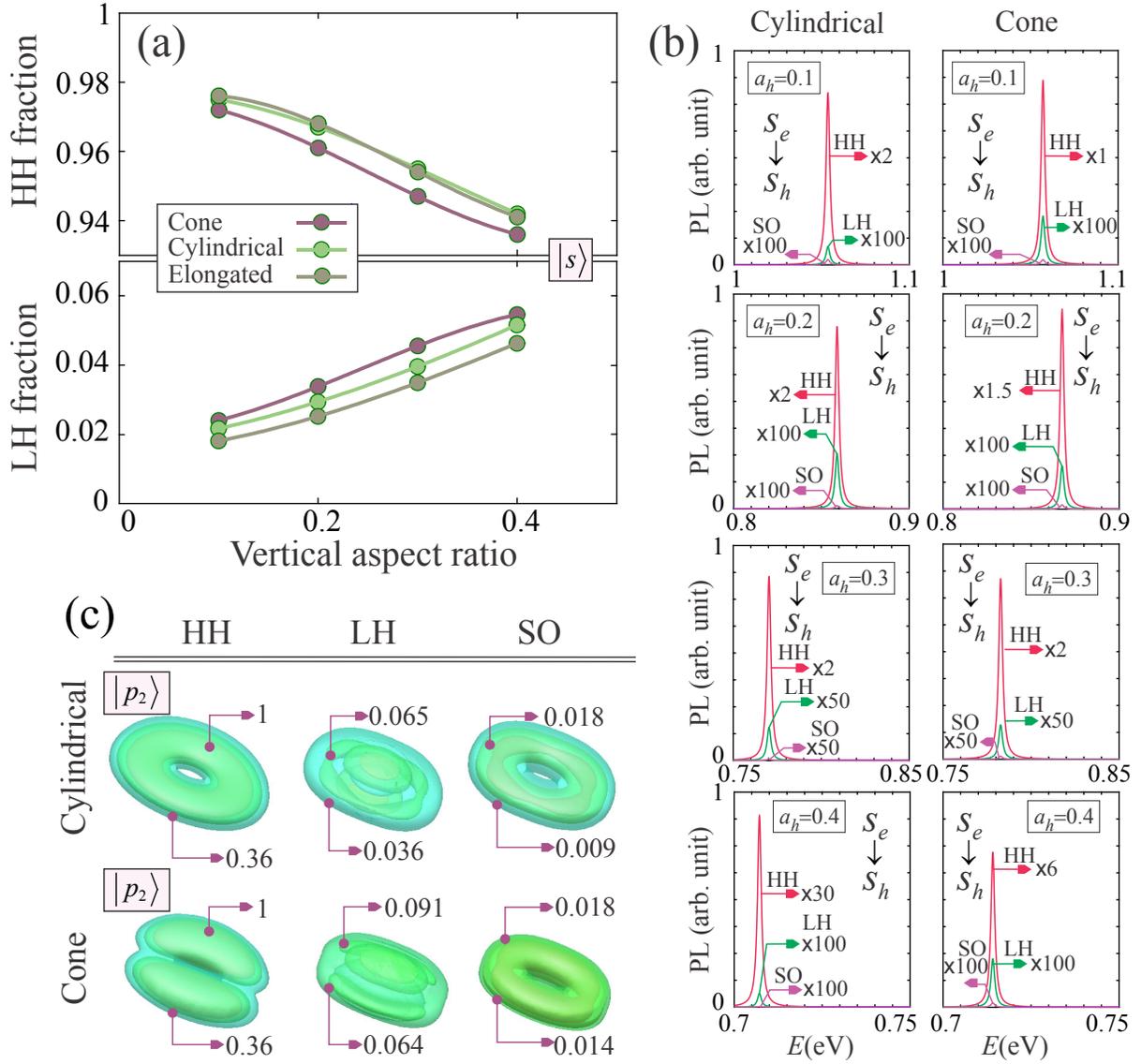

FIG. 11. HH and LH fractions in the hole ground state $|s_h\rangle$ wavefunction plotted against the $a_h$ for series B-1, C-1 and F QDs. (a) Reduction of $\Xi_{v;h}$ along with the biaxial strain redistribution enhances the LH contribution. Shear strain further increases the LH fraction in series C-1 QDs compared to their B-1 counterparts. In series F QDs, the *s*-shell hole particle exhibits an insignificant response to in-plane variations as it is quantized strongly along [001] direction. (b) HH, LH and SO non-excitonic oscillator strengths calculated for $|s_e\rangle \to |s_h\rangle$ transition in series B-1 and C-1 QDs. LH and SO oscillator strengths generally grow with $a_h$. (c) HH, LH and SO orbitals of second *p*-state $|p_2\rangle$ in series B-1 and C-1 QDs ($a_h = 0.2$). LH and SO orbitals of the hole ground state $|s_h\rangle$ are very dilute. Excited states $|p_i\rangle$ and $|d_i\rangle$ contain more SO character since they are energetically closer to the InAs bulk SO band.

Table III. Symmetry character, electron *p*-state splitting, ground state HH/LH percentages and expected FSS for NWQDs studied here, along with the same data compiled for self-assembled QDs form other theoretical results. Electron *p*-state splitting increases by enlarging the NWQD height: minimum and maximum correspond to the smallest and largest heights, respectively. Ground state HH/LH percentages are given only for $a_h$ equal to 0.2 ($h_D$ = 4 nm) in the case of NWQDs. In self assembled QDs, $h_D$ is mentioned for the available data (NA: Not Available).

| Growth vector, Shape, Material | Symmetry | | Base (nm) | Height (nm) | Electron *p*-state splitting (meV) | Ground state HH percentage | Ground state LH percentage | FSS |
|---|---|---|---|---|---|---|---|---|
| Nanowire-QD | Shape | Piezo | | | | | | |
| [001], disk, InAs/GaAs | $D_{\infty h}$ | $D_{2d}$ | 20 | 2-8 | 0.1-0.5 | 97.5 % ($h_D$ = 4 nm) | 2.17 % ($h_D$ = 4 nm) | < 1 (µeV) |
| [001], disk, InAs/GaAs | $D_{\infty h}$ | $D_{2d}$ | 25 | 4 | 0.22 | 97.9 % | 1.83 % | < 1 (µeV) |
| [001], disk, InAs/GaAs | $D_{\infty h}$ | $D_{2d}$ | 30 | 4 | 0.67 | 98.5 % | 1.32 % | < 1 (µeV) |
| [001], cone, InAs/GaAs | $C_{\infty v}$ | $C_{2v}$ | 20 | 2-8 | 0.8-9.2 | 97.2 % ($h_D$ = 4 nm) | 2.4 % ($h_D$ = 4 nm) | > 1 (µeV) |
| [001], hexagon, InAs/GaAs | $D_{6h}$ | ~$D_{2d}$ | 20 | 2-8 | 0.6-2.7 | 96.6 % ($h_D$ = 4 nm) | 2.83 % ($h_D$ = 4 nm) | > 1 (µeV) |
| [111], disk, InAs/InP | $D_{\infty h}$ | $C_{3v}$ | 20 | 2-8 | 0.4-0.6 | 95.2 % ($h_D$ = 4 nm) | 2.42 % ($h_D$ = 4 nm) | < 1 (µeV) |
| [111], cone, InAs/InP | $C_{\infty v}$ | $C_{3v}$ | 20 | 2-8 | 0.6-1.2 | 93.9 % ($h_D$ = 4 nm) | 3.03 % ($h_D$ = 4 nm) | < 1 (µeV) |
| [111], hexagon, InAs/InP | $D_{6h}$ | $C_{3v}$ | 20 | 2-8 | 0.4-0.6 | 94.6 % ($h_D$ = 4 nm) | 2.61 % ($h_D$ = 4 nm) | < 1 (µeV) |
| [001], elongated disk, InAs/GaAs | $D_{2h}$ | $C_{2v}$ | 20/30 | 2-8 | 18.4-31.6 | 96.8 % ($h_D$ = 4 nm) | 2.53 % ($h_D$ = 4 nm) | > 1 (µeV) |
| Self-assembled | | | | | | | | |
| [001], pyramid, InAs/GaAs [58] | $C_{4v}$ | $C_{2v}$ | 17 | 0.68-8.5 | 10.2 ($h_D$ = 3.6 nm) | < 93 % ($h_D$ = 0.68 nm) | 2.8-9.2 % | > 1 (µeV) |
| [001], lens, InAs/GaAs [58] | $C_{\infty v}$ | $C_{2v}$ | 17 | 2.9-8.5 | 0.3 ($h_D$ = 3.6 nm) | < 93 % ($h_D$ = 2.9 nm) | 2.8-14.4 % | > 1 (µeV) |
| [111], lens [70] | $C_{\infty v}$ | $C_{3v}$ | NA | NA | NA | NA | NA | < 1 (µeV) |
| [111], triangular, In(Ga)As/Al(Ga)As [11] | $D_{3h}$ | $C_{3v}$ | 16 | 1.5 | NA | 89 % | < 11 % | < 1 (µeV) |

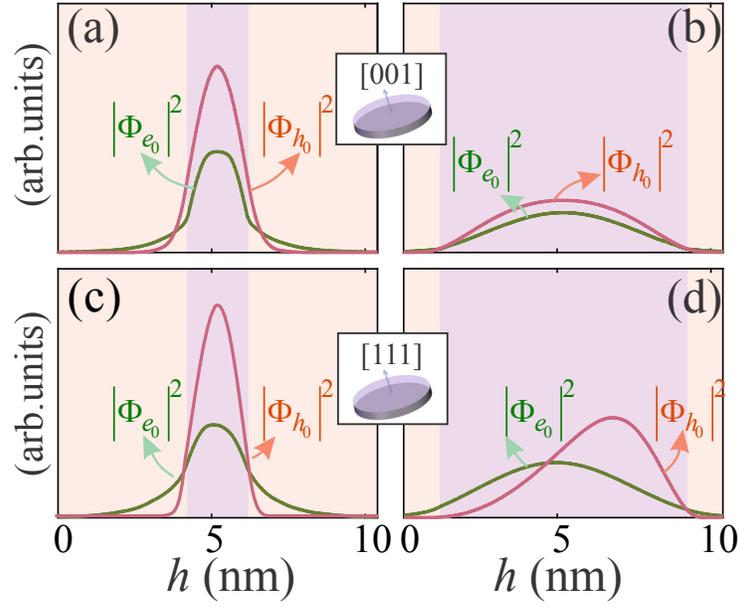

FIG. 12. Projections of the electron and hole *s*-shell orbitals on the nanowire axis: (a) and (b), series B-1 NWQD with $a_h =$ 0.1 and 0.4. The probability densities are symmetric along the axis as a consequence of trivial piezoelectric potential close to the QD center; (c) and (d), series B-2 NWQD with the same vertical aspect ratios. Electron and hole orbitals do not share same centers of mass due to the axial piezoelectric field.

Table IV. Orbital-dependent kinetic energy defined as $\langle H_k \rangle_{e_0} - E^0_{CB}$ ($\langle H_k \rangle_{h_0} - E^0_{VB}$) for electrons (holes) versus the vertical aspect ratio in different series of NWQDs.

| $a_h$ | Series B-1 | | Series B-2 | | Series C-1 | | Series C-2 | | Series D-2 | |
|---|---|---|---|---|---|---|---|---|---|---|
| | $\xi_e$ (eV) | $\xi_h$ (eV) | $\xi_e$ (eV) | $\xi_h$ (eV) | $\xi_e$ (eV) | $\xi_h$ (eV) | $\xi_e$ (eV) | $\xi_h$ (eV) | $\xi_e$ (eV) | $\xi_h$ (eV) |
| 0.1 | 0.295 | −0.056 | 0.298 | −0.057 | 0.203 | −0.071 | 0.203 | −0.073 | 0.201 | −0.073 |
| 0.2 | 0.239 | −0.040 | 0.245 | −0.041 | 0.198 | −0.045 | 0.202 | −0.047 | 0.195 | −0.045 |
| 0.3 | 0.201 | −0.030 | 0.213 | −0.031 | 0.180 | −0.037 | 0.188 | −0.038 | 0.175 | −0.035 |
| 0.4 | 0.177 | −0.025 | 0.199 | −0.030 | 0.163 | −0.033 | 0.179 | −0.036 | 0.159 | −0.032 |
| 0.5 | 0.162 | −0.028 | 0.198 | −0.034 | 0.151 | −0.036 | 0.176 | −0.041 | 0.147 | −0.044 |

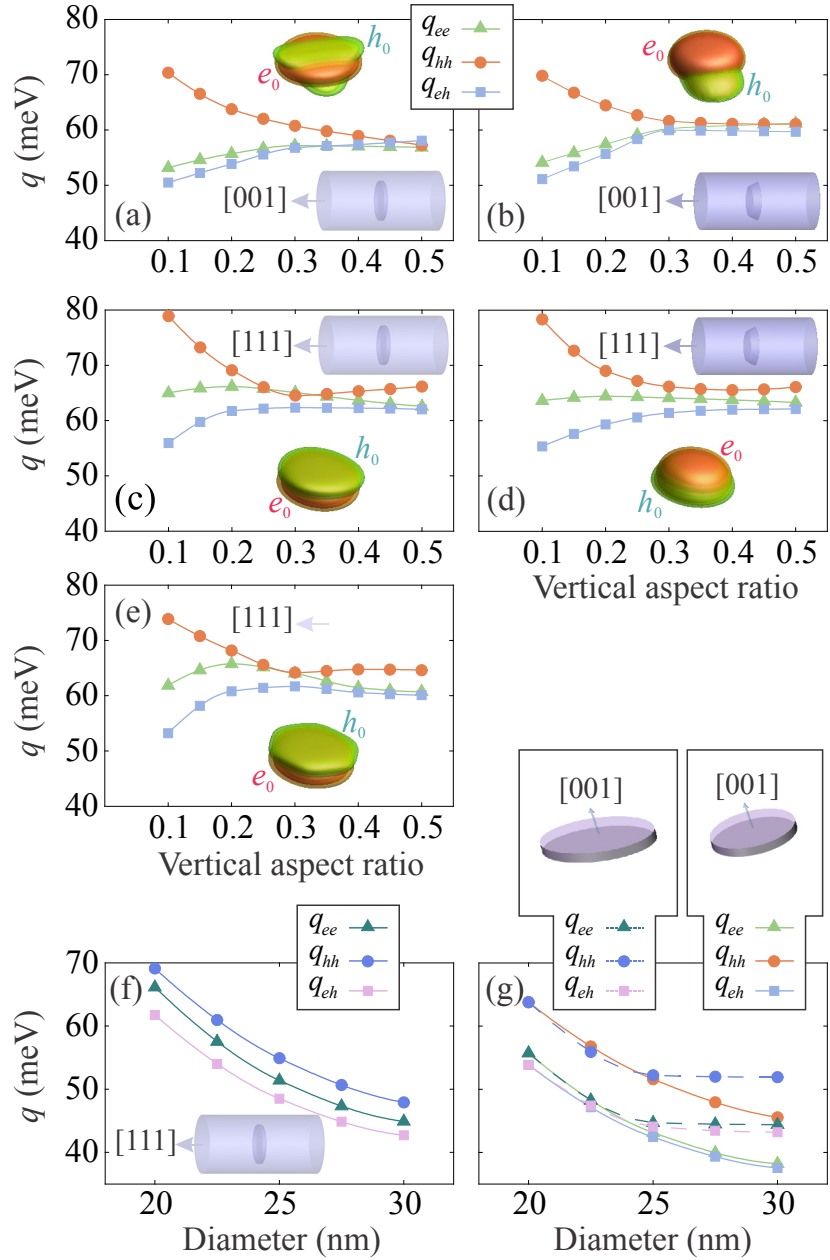

FIG. 13. Direct Coulomb interaction terms for the ground state electrons and holes in different series of NWQDs. Temporary increase in $q_{00}^{d,ee}$ and $q_{00}^{d,eh}$ results from the reversed penetration of electron orbital from the nanowire core into the insertion region. Insets show the $s$-shell electron and hole respective positions for $a_h = 0.4$ where $q_{00}^{d,ee}$, $q_{00}^{d,hh}$ and $q_{00}^{d,eh}$ become comparable. (f) and (g): Enlarging the QD diameter pronouncedly decreases the Coulomb terms in a similar fashion ($\propto 1/D_D$); QD height $h_D$ is kept fixed equal to 4 nm. (g) [001]-oriented NWQDs are elongated in [100] direction less than 50%. Electron and hole hardly spread laterally having $D_2/D_1 > 1.25$, thus direct Coulomb terms approximately remain fixed.

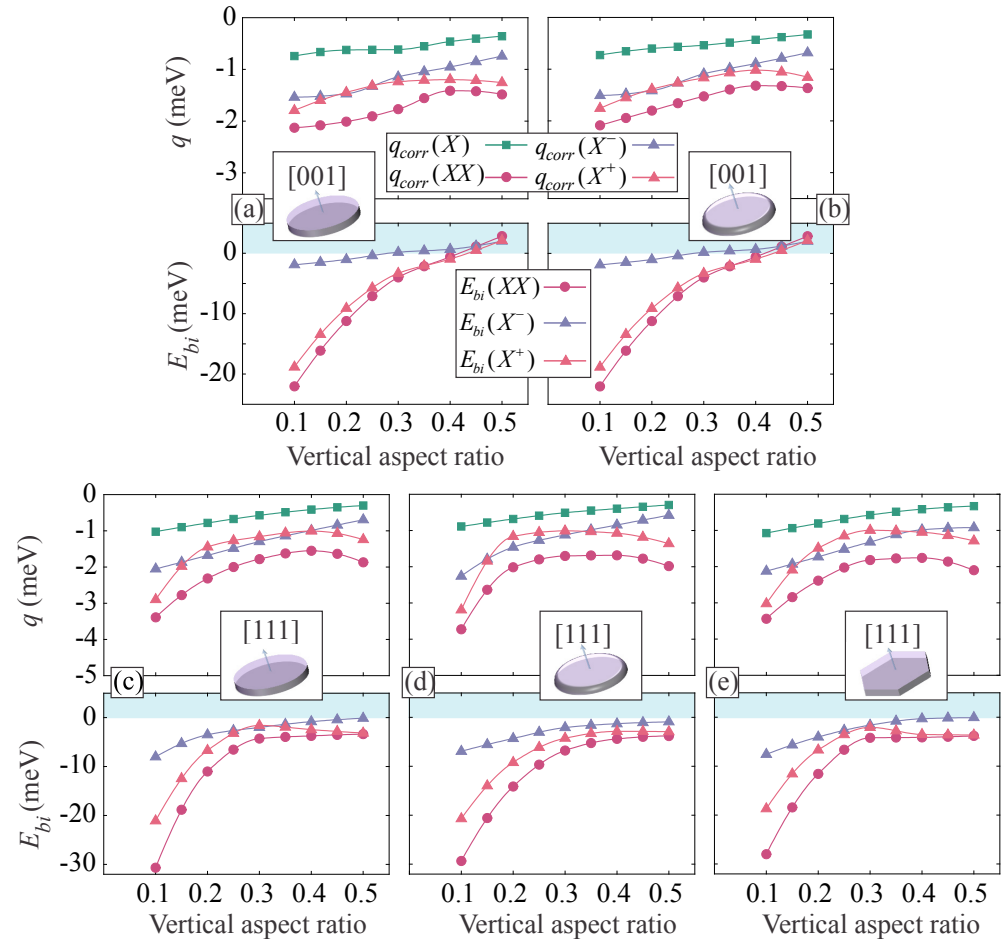

FIG. 14. Coulomb correlation and binding energies in different series of NWQDs plotted for *s*-shell $X$, $XX$, $X_0^+$ and $X_0^-$. Correlation energies are larger wherever the repulsive interaction terms significantly exceed the attractive term. They also grow for larger $a_h$ where the chance of exchanging kinetic energy becomes smaller.

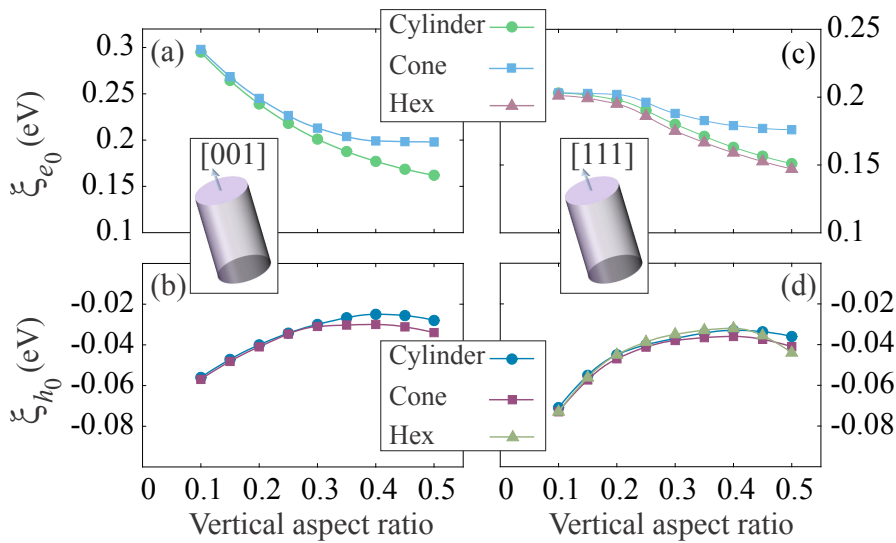

FIG. 15. Orbital-dependent kinetic energy of electrons $\xi_{e_0}$ and holes $\xi_{h_0}$ plotted for [001]-oriented (a, b) and [111]-oriented (c, d) nanowire QDs. No remarkable distinction exists for flat QDs as they possess similar orbital spreadout and hole LH character. For larger QD heights, electrons and holes in cylindrical type gain the least $\xi_{e_0;h_0}$ having the smaller orbital extent (and LH contribution for holes). Dispersion is steeper in cone-shaped QDs, plus that hole orbital is pushed downward occupying larger area despite the stronger piezoelectric field and smaller QD size.

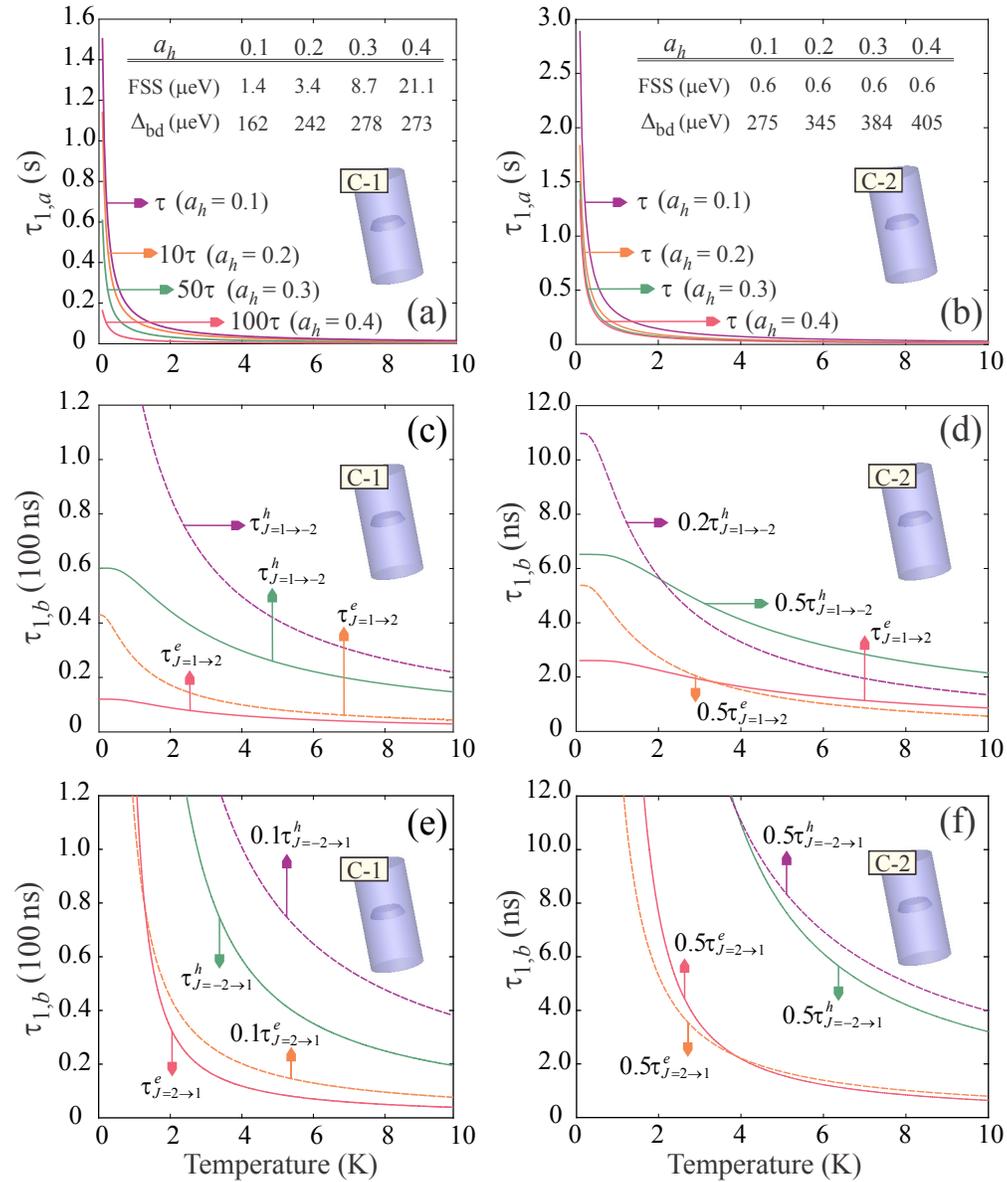

FIG. 16. Relaxation times, due to the first ($\tau_{1,a}$) and second ($\tau_{1,b}$) spin flip processes introduced in Sec. III. B, are compared for [001] and [111]-oriented cone shaped NWQDs. In the first mechanism, both exchange energies $\delta_s$ (FSS) and $\Delta_{bd}$ manipulate the relaxation time. (a) $\tau_{1,a}$ rapidly drops for larger aspect ratios since both $\Delta_{bd}$ and $\delta_s$ grow with $a_h$ in series C-1 QDs ($\tau_{1,a} \gg$ exciton lifetime $\tau_X$) (b) $\tau_{1,a}$ is plotted for maximum $\delta_s$ in our calculations. $C_{3v}$ symmetry character of exciton in series C-2 QDs is independent of $a_h$, thus $\delta_s$ is ideally zero and only $\Delta_{bd}$ tunes the relaxation time ($\tau_{1,a} \gg \tau_X$). (c) and (d) $\tau_{1,b}$ ($X_B \rightarrow X_D$) versus temperature calculated for series C-1 and C-2 QDs with two different diameters, 20 (solid line) and 30 nm (dashed line), and $h_D$ = 4 nm. (d) and (f) $\tau_{1,b}$ ($X_D \rightarrow X_B$) for the same structures in (c) and (d), respectively.

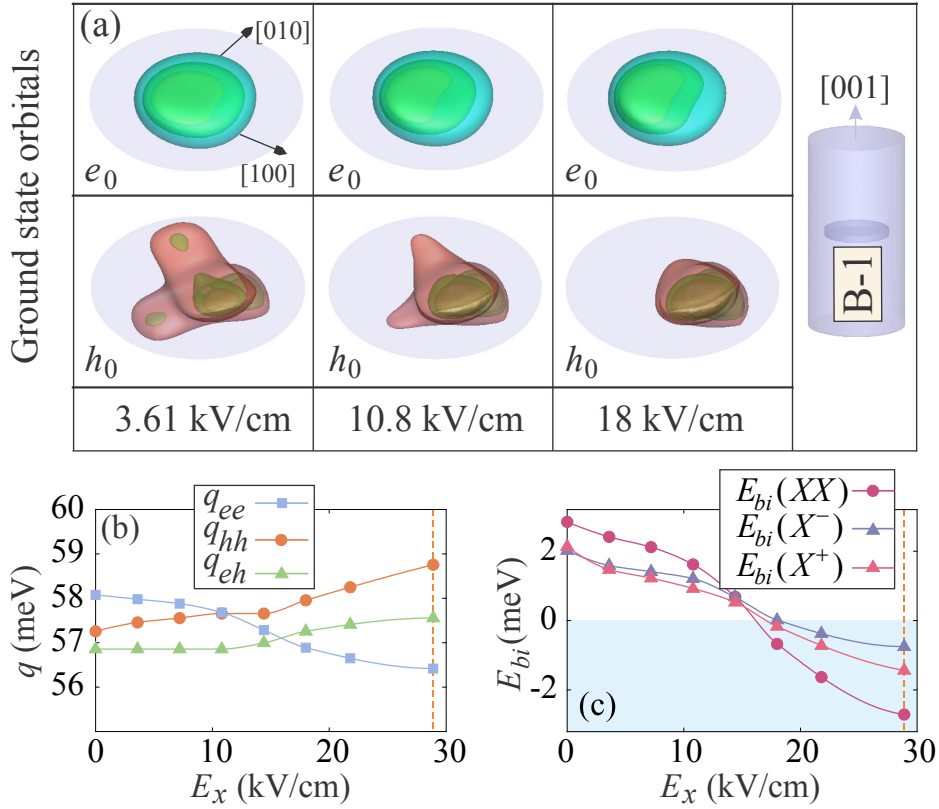

FIG. 17. (a) Hole and electron orbitals respond to $E_x \parallel [100]$ by squeezing their associated orbitals. Effective field amplitudes $E_{\text{eff}}$ indicated above correspond to 25, 75 and 125 kV/cm external electric fields respectively. The destruction of $D_{2d}$ symmetry gives rise to noticeable anisotropic exchange interaction and FSS. Hole orbital flexibly moves due to the small cost of kinetic energy it spends by reshaping. (b) Direct Coulomb interactions versus $E_x$. Range of variations due to the electric field is quite trivial compared to the variations in response to $\Xi_{v;e,h}$. (c) Ground-state complexes show a binding to antibinding transition. Correlation energies are barely affected thus binding energies follow the evolution of direct interaction terms.

Table V. HH and LH percentages of the ground state orbital in series B-1 NWQDs against [001]-oriented external electric field.

| $E_{[100]}$ (kV/cm) | HH | LH |
|---|---|---|
| 0 | 92.6 | 6.59 |
| 25 | 91.9 | 5.62 |
| 75 | 91.0 | 7.20 |
| 125 | 92.6 | 8.02 |

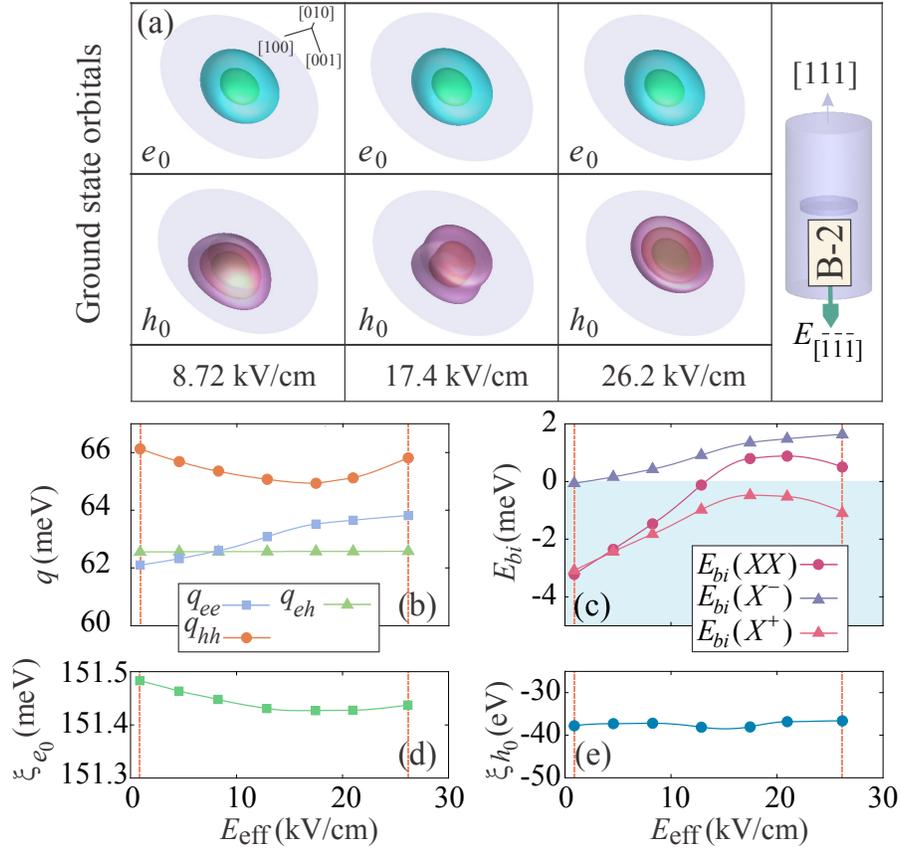

FIG. 18. (a) Electron and hole orbitals in the presence of an axial electric field $E_{ex}$ in [111]-oriented QDs. Hole is well-confined beneath the top interface when the parallel component of piezoelectric field dominates the effective applied field ($E_{eff}$ = 8.72 kV/cm). Hole orbital moves downward as a function of effective electric field until reaching the bottom interface. Electron orbital, however, stands unmoved. (b) Direct Coulomb interactions are shown against $E_{eff}$: $e_0$ and $h_0$ spatial overlap is improved while $q_{00}^{hh}$ undergoes a transient decrease meanwhile $h_0$ orbital spreads vertically. (c) Binding energies of the ground-state complexes. Biexciton becomes weakly binding at $E_{eff} \approx 12$ kV/cm. (d) and (e): hole and electron orbital-dependent kinetic energies versus the axial electric field.